\documentclass[12pt,preprint]{aastex}
\usepackage{graphicx}

 \makeatother
\def\lcdm  {\rm $\Lambda$CDM\ }

\def\Mpc{{\rm\,Mpc}}
\def\mpc{{\rm\,Mpc}}

\def\msun{{\rm\,M_\odot}}

\def\vol#1  {{{#1}{\rm,}\ }}

\def\etal{et al.\ }
\def\sec{\S }

\def\geff{\gamma_{\rm eff}}

\newcount\refno
\newcount\rfno
\def\eq{$^{\the\refno\ }$\advance\refno by 1}
\def\ad{\advance\rfno by 1}

\def\clock{\count0=\time \divide\count0 by 60
     \count1=\count0 \multiply\count1 by -60 \advance\count1 by \time
     \number\count0:\ifnum\count1<10{0\number\count1}\else\number\count1\fi}

\def\myputfigure#1#2#3#4#5%
{\hskip0.03\textwidth\vskip#5pt%\hfill
\makebox[0pt]{\hskip#2in
\includegraphics[width=#3\textwidth]{#1}}\vskip#4pt\hfill}

\def\Gcm2{\rm G~cm^2}

\def\beq{\begin{equation}}
\def\eeq{\end{equation}}
\def\bea{\begin{eqnarray}}
\def\eea{\end{eqnarray}}

\def \date         {\ifcase\month \message{zero} \or
                    January \or February \or March \or April \or May \or June
                    \or July \or
                    August \or September \or October \or November \or
                    December \fi
                    \space\number\day, \number\year}

\def\kev{{\rm keV}}

\def\cmb{_{{\rm CMB}}}

\def\om{\Omega_m}
\def\ob{\Omega_b}
\def\ol{\Omega_{\rm \Lambda}}
\def\mvir{M_{\rm vir}}
\def\rvir{r_{\rm vir}}
\def\tvir{T_{\rm vir}}
\def\pix{{\rm pixel}}
\def\c{c_{\rm vir}}
\begin{document}

\title{WMAP constraints on the Intra-Cluster Medium}
\author{Niayesh Afshordi\altaffilmark{1,2}, Yen-Ting Lin\altaffilmark{3}, and Alastair
J.R. Sanderson\altaffilmark{3,4}} \altaffiltext{1}{Institute for
Theory and Computation, Harvard-Smithsonian Center for Astrophysics,
MS-51, 60 Garden Street, Cambridge, MA 02138;
nafshordi@cfa.harvard.edu} \altaffiltext{2}{Princeton University
Observatory, Princeton University, Princeton, NJ 08544}
\altaffiltext{3}{Department of Astronomy, University of Illinois,
Urbana, IL 61801} \altaffiltext{4}{School of Physics and Astronomy,
University of Birmingham, Edgbaston, Birmingham B15 2TT, UK}
% \accepted{ }

\begin{abstract}
    We devise a Monte-Carlo based, optimized filter match method to extract the thermal
    Sunyaev-Zel'dovich (SZ) signature of a catalog of 116 low-redshift X-ray
    clusters from the first year data release of the Wilkinson Microwave
    Anisotropy Probe (WMAP). We detect an over-all amplitude for the SZ signal at the $\sim 8\sigma$
    level, yielding a combined constraint of $f_{\rm gas}h = 0.08 \pm 0.01 (ran) \pm 0.01 (sys)$ on the gas
    mass fraction of the Intra-Cluster Medium. We also compile X-ray
    estimated gas fractions from the literature for our sample, and
% NOTE: see -> find
    find that they are consistent with the SZ estimates at the
    $2\sigma$ level, while both show an increasing trend with X-ray temperature.
     Nevertheless, our SZ estimated gas fraction is
    $30-40\%$ smaller than the concordance \lcdm cosmic average. We
    also express our observations in terms of the SZ flux-temperature
    relation, and compare it with other observations, as well as
    numerical studies.

    Based on its spectral and spatial signature, we can also
    extract the microwave point source signal of the clusters at the
    $3\sigma$ level, which puts the average microwave luminosity (at $\sim 41$ GHz) of
% NOTE: < -> \le
    bright cluster members ($M_K \le -21$) at $(2.4 \pm 0.8) \times 10^{27}
    h^{-2} {\rm erg/s/Hz}$ . Furthermore, we can constrain the average dark matter halo concentration
    parameter to $c_{\rm vir}=3.4^{+0.6}_{-0.9}$, for clusters with $T_{\rm x} > 5~\kev$.

    Our work serves as an example for how correlation of SZ surveys
    with cluster surveys in other frequencies can significantly
    increase our physical understanding of the intra-cluster medium.

\end{abstract}

\section{Introduction}

Clusters of galaxies are the largest relaxed concentrations of mass
in the universe. They are interesting for cosmology as they probe
% NOTE: wording change, ref added
the evolution of the large scale structure of the universe
\citep[e.g.][]{eke96,viana99,Haimanetal2001,HuHaiman2003}
%the large scale structure of the universe \citep[e.g.,][]{Mohretal2002, HuHaiman2003},
and they are interesting on their own, as we can resolve, probe, and study their
inner structure in different frequencies, ranging from microwave to
X-rays, and also through weak and strong gravitational lensing of
background galaxies
\citep[e.g.,][]{carlstrom,nichol,markev,hennawi,massey,sand}. What
adds to this simplicity is that, at least for the massive clusters,
almost all of the baryonic matter sits in the diffuse ionized
Intra-Cluster Medium (ICM), which can be studied theoretically and
% NOTE: comma deleted
observationally with relatively simple physics, and give us a
census of cosmic baryonic budget
% NOTE: ref changed
\citep[see e.g.,][]{Whiteetal,Evrard,mohr99}.

In this paper, we focus on the microwave signatures of galaxy
clusters, the thermal Sunyaev-Zel'dovich (SZ) effect \citep{SZ},
caused by the scattering of Cosmic Microwave Background (CMB)
photons by hot gas in the diffuse ICM, and yielding characteristic
spatial and spectral imprints on the CMB sky.

 The thermal SZ effect has changed from the subject of theoretical
 studies to that of intense observational endeavor within the past decade,
 as various experiments have and
are being designed to study this effect \citep[e.g., APEX, ACT, AMI,
Planck, SZA, SPT; see][for an overview]{bond}. The main reason
behind this wide attention is the potential for using SZ detected
clusters as standard candles to probe the cosmological evolution up
to large redshifts
%  NOTE: add ref lin03b,majumdar03
\citep[e.g.,][]{Haimanetal2001,verde,lin03b,majumdar03}.
Compared to the SZ detection
method, the X-ray detected clusters, which have been primarily used
for this purpose until now \citep[e.g.,][]{henry1,vikhlinin,henry2},
become much harder to detect at large redshifts, and are also
believed to be more affected by complex astrophysics associated with
galaxy formation, cooling or feedback within clusters
\citep{carlstrom}.

Moreover, unlike X-ray observations which only sample regions of
high gas density, thermal SZ observations probe the distribution of
thermal energy in the cluster, and thus provide independent
information about the over-all thermal history \citep[Is there an
entropy floor?; e.g.,][]{voit2002,ponman}, and baryonic budget of
the cluster \citep[Are there missing baryons?;
e.g.,][]{cenostriker}.

Although various scaling relations of X-ray properties of clusters
have been extensively studied in the literature, mainly due to the
scarcity, incoherence, or low sensitivity of SZ observations of
clusters, there have been only a few statistical analyses of SZ
scaling properties in the literature \citep{cooray,mccarthy,benson}.
Given the upcoming influx of SZ selected cluster catalogs, a good
understanding of these scaling relations, and in particular, the SZ
flux-Mass relation (see \S \ref{SZprofile}), which is of special
significance for cosmological interpretations of these catalogs
% NOTE: ref added
(e.g. \citealt{majumdar03}), is
still lacking.

%%The first year data release of the Wilkinson Microwave Anisotropy
%%Probe \citep[WMAP;][]{WMAP} has constrained our cosmology with an
%unprecedented accuracy. Most of these constraints come from the
%linear fossils of the early universe which have been preserved in
%the temperature anisotropies of the Cosmic Microwave Background
%(CMB). These are the ones that can be easily understood and dealt
%with, within the framework of linear perturbation theory. However,
%there are also imprints of the late universe which could be seen in
%the WMAP results. Most notably, the measurement of the optical depth
%to the surface of last scattering, $\tau\simeq 0.17$, which implied
%an early reionization of the universe, was the biggest surprise.
%There is also the strangely small amplitude of the large-angle CMB
%anisotropies which remains unexplained \citep{Spergel2003}.

While the first year data release of the Wilkinson Microwave
Anisotropy Probe \citep[WMAP;][]{WMAP} has constrained our cosmology
with an unprecedented accuracy, due to its low resolution and low
frequency coverage, the SZ effect
 cannot be directly observed in the WMAP CMB maps \citep{Huffenberger}.
 One possible avenue is cross-correlating CMB anisotropies with a tracer of the density
% NOTE: trace -> traces
 (which traces clusters and thus SZ signal) in the late universe \citep{Peiris,ZhangPen}.
In fact, different groups have reported a signature of
anti-correlation (which is what one expects from thermal SZ at WMAP
frequencies) at small angles between WMAP maps and different galaxy
or cluster catalogs, at a few sigma level
\citep{WMAPfor,fosalba1,fosalba2,Myers2004,ALS}. While thermal SZ is
the clear interpretation of this signal, relating such observations
to interesting cluster properties can be confused by the physics of
non-linear clustering or galaxy bias.

\citet{hernand1} and \citet{hernand2} use an alternative method,
where they construct SZ templates based on given cluster or galaxy
catalogs, and then calculate the over-all amplitude of WMAP signal
temperature decrement associated with that template. While the
method yields significant SZ detections (2-5$\sigma$), the physical
interpretation is complicated by the non-trivial procedure that they
use to construct these templates.

  In this paper, we follow the second line by devising an optimized filter match method
based on an analytic model of ICM which is motivated by both
numerical simulations and observations. We then apply the method to
a sample of X-ray clusters, and construct templates of both SZ and
potential point source contamination based on the X-ray temperatures
of each cluster. Combining these templates with the WMAP maps yields
constraints on the physical properties of our ICM model, namely the
ICM gas mass fraction, and the dark matter halo concentration
parameter.

% NOTE: wording change
Almost all the SZ observations up to date use an isothermal $\beta$ model,
obtained from X-ray observations, to describe the cluster SZ profile.
However, it has been
%    While almost all the SZ observations up to date use an
%    isothermal $\beta$ model fit to the X-ray observations, in
%    order to obtain the cluster SZ profile, it has been
    demonstrated that, as X-rays and SZ cover different scales
    inside the cluster, such extrapolation can lead to errors as
    big as a factor of $2$ in the interpretation of SZ observations \citep{schmidt}.
    Instead, \citet{schmidt} suggest using a physically motivated
    NFW profile (see \sec \ref{model}) to model the X-ray and SZ
    observations simultaneously, which in their case, leads to consistent estimates of Hubble constant for
     three different clusters.
We choose to follow their approach in
     choosing a physically motivated ICM model, rather than a
     mathematically convenient one.

    In Appendix A, we introduce a semi-analytic NFW-based model for the ICM gas profile.
    We then start in \sec \ref{data} by describing the WMAP CMB
    temperature maps and our compiled X-ray cluster catalog.
    \sec \ref{model} derives the theoretical SZ/Point Source templates based on our ICM model, while \sec
    \ref{CMBstat} describes our statistical analysis methodology.
    In \sec \ref{results} we describe the results of our statistical
    analysis, listing the constraints on gas fraction, concentration
    parameter, and point source contamination of our clusters.
    It the end, \sec \ref{discuss} discusses the validity of various
    assumptions that we made through the treatment, and \sec
    \ref{conclusions} highlights the major results and concludes the
    paper.

    Throughout the paper, we assume a \lcdm flat cosmology with $\om
    = 0.3$, and $H_0 = 100 h ~{\rm km/s/Mpc}$. While no assumption for $h$ is made in our analysis of the SZ signal,
     we adopt the value of $h=0.7$ to compare the X-ray gas
     fractions with our SZ signal.

\section{Data}\label{data}
\subsection{WMAP foreground cleaned CMB maps}\label{WMAP}

We use the first year data release of the observed CMB sky by WMAP
for our analysis \citep{WMAP}. The WMAP experiment observes the
microwave
  sky in 5 frequency bands ranging from 23 to 94 GHz. The detector
  resolution increases monotonically from 0.88 degree for the lowest
  frequency band to 0.22 degree for the highest frequency. Due
  to their low resolution and large Galactic contamination, the two bands with the lowest
  frequencies, K(23 GHz) and Ka(33 GHz), are mainly used for Galactic foreground
  subtraction and Galactic mask construction \citep{WMAPfor},
  while the three higher frequency bands, which have the highest
  resolution and lowest foreground contamination, Q(41 GHz), V(61 GHz), and
  W(94 GHz), are used for CMB anisotropy spectrum analysis \citep{WMAPpower}.
   \citet{WMAPfor} combine the frequency dependence of 5 WMAP bands with
  the known distribution of different Galactic components that
  trace the dominant foregrounds (i.e. synchrotron, free-free, and dust emission)
  to obtain the foreground contamination in each band. This foreground map is then used to
  clean the Q, V and W bands for the angular power spectrum
  analysis. Similarly, we use the cleaned temperature maps of
  these three bands for our SZ analysis. We also
  use the same sky mask that they use, the Kp2 mask which masks out 15\% of the sky,
  in order to avoid any remaining Galactic foreground. However, we {\it stop short of}
   masking out the 208 identified WMAP point sources, as many of them happen to be close
   to our clusters. For example, there are 29 WMAP identified microwave sources within
   3.6 degrees of 66 of our clusters.
% NOTE: slight wording change
Instead, we decide to model the point source contamination based on its
frequency dependence (\sec \ref{secps}).

% NOTE: add "the"
    The WMAP temperature maps and mask are available in the HEALPix
    format of spherical coordinate system \citep{healpix}, which is an equal-area
    pixelization of the sphere. The resolution of the first year
    data is $N_{side}=512=2^9$, implying $12\times N_{side}^2 =
    3,145,728$ independent data points (in lieu of masks) and $\simeq 0.1^{\circ}$ sized pixels,
     for each sky map.

\subsection{Cluster Catalog and X-ray Data}\label{catalog}

Our objective is to study the SZ signal in a large sample of galaxy
clusters. To this end, we have assembled our sample from several
existing X-ray cluster samples
% NOTE: add ref, change order
\citep[][]{david93,mohr99,jones99,FRB,reiprich02,ikebe,sanderson03}, as
%\citep[][]{mohr99,reiprich02,FRB,sanderson03,david93,jones99,ikebe}, as
X-ray observations may provide reliable cluster mass estimates, and
avoid false detections due to chance projections. The selection
criteria require that the clusters (1) must have {\it measured}
X-ray emission weighted mean temperature ($T_{\rm x}$), (2) are
reasonably away from the Galactic plane (Galactic latitude
$|b|>10^\circ$), and (3) are at least 3 degrees away from
the Galactic foreground Kp2 mask (\sec \ref{WMAP}).%, and (4) have
%reliable ICM gas mass estimates (see below).

The redshift information is obtained from NED and/or SIMBAD, and the
% NOTE: wording change
above catalogs. The cluster temperature is taken from the literature cited
above, primarily the study of \citet{ikebe}. We have
adopted the $T_{\rm x}$ obtained when the central cool core region
% NOTE: wording change
is excluded, and identify the peak of the X-ray
emission (either from the cluster catalogs or from archival {\it
ROSAT} images) as the cluster center. Our final cluster catalog contains 117 nearby
clusters, ($0 \lesssim z \lesssim 0.18$), whose temperature ranges
from 0.7 to 11 keV.

Our requirement that clusters have measured $T_{\rm x}$ is to
provide reliable mass estimates. Given $T_{\rm x}$, the observed
mass-temperature relation \citep[][hereafter FRB01]{FRB}
\begin{equation}
M_{500} = \left( 1.78^{+0.20}_{-0.17}\,\times 10^{13} h^{-1}
M_{\odot}\right)
    T_{\rm x}(\kev)^{1.58^{+0.06}_{-0.07}}
\end{equation}
can be used to obtain $M_{500} \equiv (4\pi/3) 500 \rho_c
r_{500}^3$, the mass enclosed by $r_{500}$, within which the mean
overdensity is 500 times the {\it critical density } of the universe
$\rho_c$. The wide range of cluster temperature in our sample
implies that our clusters span two orders of magnitude in mass.

In order to facilitate comparisons of our SZ-derived gas fraction
with the X-ray measurements, we compile the gas mass (within
$r_{500}$) for most of our clusters from the literature
\citep{mohr99,jones99}, supplemented by the data based on the study
of \citet{sanderson03}. The gas masses provided by \citet{jones99}
are measured at a fixed metric radius of $0.5 h^{-1}$ Mpc; we
convert it to the nominal radius of $r_{500}$ by the
% NOTE: remove "(or estimated)", add "the virial radius"
measured $\beta$-model profile, and then to the virial radius $r_{200}$
($=\rvir$; see \sec \ref{model}), using the analytic model of \sec
\ref{model}.

\begin{figure}
\includegraphics[width=\linewidth]{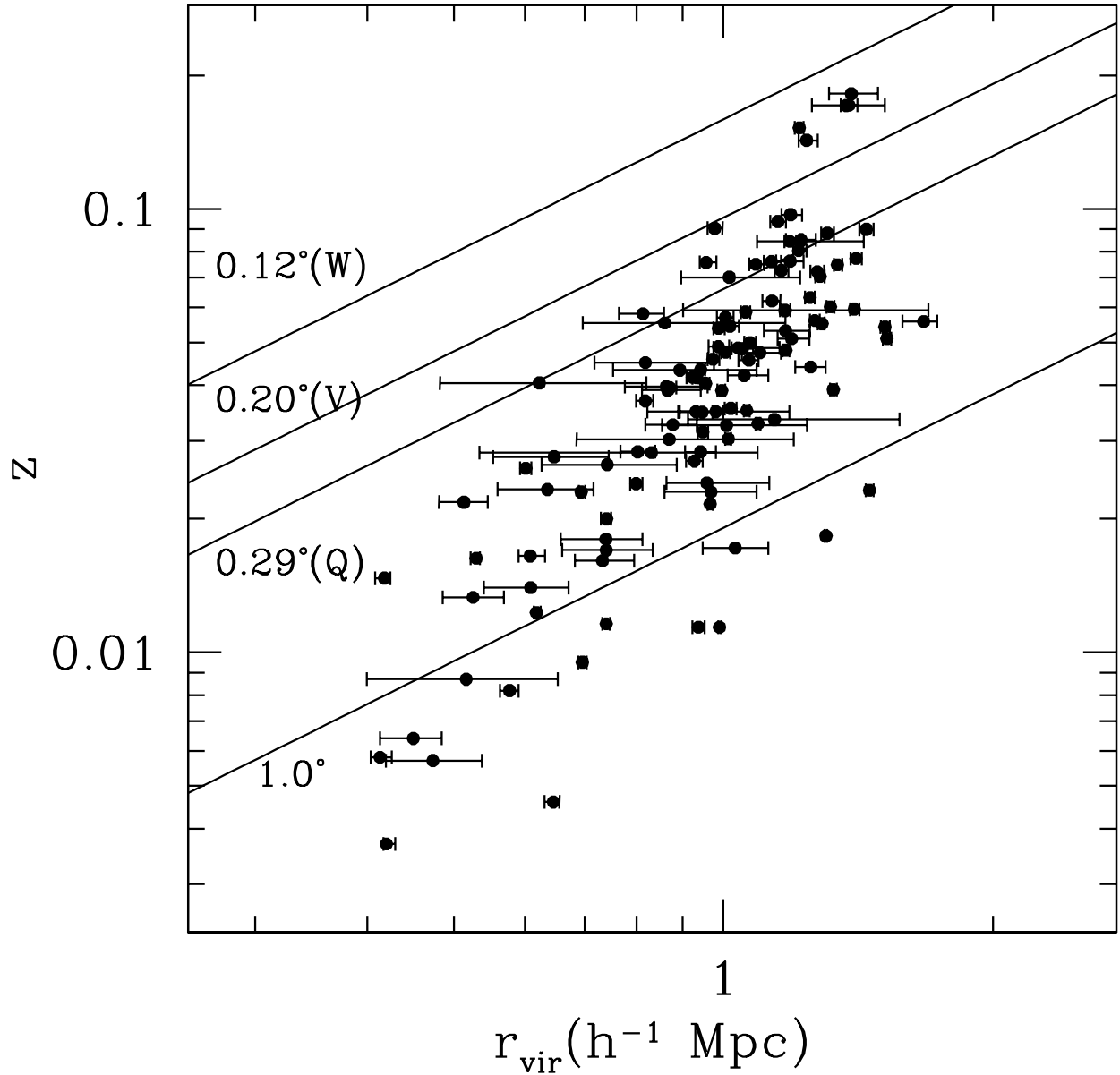}
\caption{\label{zrvir} The distribution of cluster redshifts and
virial radii (estimated from X-ray temperature, assuming $\c=5$; see
\sec \ref{model}). The three upper lines show the resolution of WMAP
bands \citep[associated with the radius of the disk with the same
effective area as the detector beams; see][]{WMAPbeam}, while the
lower line shows the physical radius of the 1 degree circle
 at the cluster redshift. }
\end{figure}

 Fig. (\ref{zrvir}) shows the distribution of redshifts and virial
 radii for our clusters (using the analytic model of \sec \ref{model}
% NOTE: add "concentration"
 for dark matter concentration $\c=5$). The solid lines show the resolution
 limits of the three
 WMAP bands, as well as the physical radius of the 1 degree circle
 at the cluster redshift. We see that most of our clusters are in fact resolved
 in all the WMAP bands.

\section{Modeling the Intra-Cluster Medium}\label{model}

         In Appendix A, based on the assumption of hydrostatic
         equilibrium and NFW dark matter profile \citep{NFW}, we
         develop an analytic model for the gas and temperature
         distribution in the Intra-Cluster Medium (ICM). In this
         model, assuming a given NFW concentration parameter,
         $\c$, all the properties of the cluster/ICM are
         quantified in terms one parameter, which, can be taken to be
         e.g., the cluster virial mass, $\mvir$, or its X-ray temperature
         $T_{\rm x}$ (see Equations \ref{delta} and \ref{tx} for
         definitions). In particular, $\mvir$ can be
         expressed in terms of (the observed) $T_{\rm x}$ (Equation \ref{tx})\footnote{Note that we have assumed the observed X-ray
         temperature, $T_{\rm x}$, to be the emission weighted temperature in our model. We address the error introduced due to
         this assumption in \sec\ref{model_unc}.} within the model for a given value of
         $\c$. Now, let us estimate the dominant microwave signals of a
         galaxy cluster, based on our simple model.

\subsection{SZ profile}\label{SZprofile}

        The contribution of the thermal SZ effect to the CMB
        temperature anisotropy \citep[see][for a review]{carlstrom}, at the frequency $\nu$,
        is proportional to the integral of electron pressure along
        the line of sight
        \beq
        \delta T_{SZ}(\nu) = -\frac{\sigma_T~ T_{\cmb} F(h\nu /T_{\cmb})
        }{m_e c^2} \int P_e ~dr, {\rm ~ where~~} F(x) = 4
        -x\coth(x/2),
        \eeq
        where $\sigma_T$ is the Thomson scattering cross-section,
        and  $m_e$ is the electron mass. The SZ flux, defined as the
        integral of $\delta T_{SZ}$ over the solid angle, $\delta \Omega$, is then
        given by
        \beq
        \overline{\delta T}_{SZ} \cdot \delta \Omega \equiv \int_{\delta \Omega} \delta T_{SZ} (\nu; {\bf \hat{n}})
        ~d^2{\bf \hat{n}}
        = -\frac{\sigma_T T_{\cmb} F(h\nu /T_{\cmb})}{m_e c^2}
        \int_{\delta \Omega} P_e({\bf x})\cdot d^{-2}_A({\bf x})~ d^3{\bf
        x}.
        \eeq
        Here, $d_A$ is the angular diameter distance, and ${\bf x}$
        spans over the cone extended by the solid angle $\delta
        \Omega$. Now, assuming local thermal equilibrium
        \beq
        P_e = \left(2+2X\over 3+5X\right) P_g \simeq 0.52 ~P_g,
        \eeq
        the total SZ flux of a cluster is
        \bea
         S_{tot}(\nu) = \overline{\delta T}_{SZ} \cdot \delta \Omega |_{tot} = -\frac{(1+X)~
        \sigma_T ~ T_{\cmb}~F(h\nu /T_{\cmb})~ f_{\rm gas} ~\mvir ~\tvir}{2 ~m_p~ m_e c^2~
        d^2_A}\nonumber\\
        = - (1.42 \times 10^{-2} ~{\rm mK})\left( 0.1 ~{\rm deg} \right)^2 \left\{F(h\nu /T_{\cmb})~ f_{\rm gas} ~
        h \over \left[H_0 d_A(z)/c\right]^2\right\} \tvir (\kev) \left(\mvir \over
         10^{15} h^{-1} \msun \right),
        \eea
         which can be combined with our analytic model (equations \ref{rvir}, \ref{tx}, and \ref{tvir}) to
         obtain
         \bea
        S_{tot}(\nu)= - (2.16 \times 10^{-4} ~{\rm mK})\left( 0.1 ~{\rm deg} \right)^2
        ~\left\{F(x)\over E(z) \left[H_0 d_A(z)/c\right]^2\right\} T^{5/2}_{\rm x} (\kev) \left[B(\c)~ f_{\rm gas} ~
        h \right] \nonumber\\
            = - (1.63 \times 10^{-2} ~{\rm mJy})
        ~\left\{x^4 F(x)\over \sinh^2 (x/2) E(z) \left[H_0 d_A(z)/c\right]^2\right\} T^{5/2}_{\rm x} (\kev) \left[B(\c)~ f_{\rm gas} ~
        h \right] \label{szflux}
         \eea
         where
         \bea
          x=h\nu /T_{\cmb}, \\
          B(\c) = \left(\int f ~g ~x^2dx \over \int g~x^2dx
          \right)\cdot \left(\int f^{3/2}~g^2~x^2dx \over \int
          f^{1/2}~g^2~x^2dx \right)^{-5/2}, \\
          E(z) = H(z)/H_0 = \left[\om (1+z)^3 + \ol \right]^{1/2},
          \\
          {\rm }~~ H_0d_A(z)/c = (1+z)^{-1} \int_0^z
          \frac{dz^{\prime}}{E(z^{\prime})},
        \eea
% NOTE: add a sentence
         and functions $f(x)$ and $g(x)$ are defined in Appendix A (Equations \ref{fdef} and \ref{gdef}). For the relevant range of $3<\c<8$, which is consistent with various
measurements of cluster dark matter profile (see \citealt{lin04} for a brief
review), $B(\c)$ is a decreasing function of $\c$ which varies from
         $2.1$ to $1.4$. Note that all the factors in equation
         (\ref{szflux}), with the exception of the last one, are fixed by
         observations. Therefore, $\left[B(\c)f_{\rm gas}h\right]$ is the
         combination of model parameters which will be fixed by our
         SZ flux observations.

% NOTE: remove comma
         We then use a Monte-Carlo method to reproduce the expected SZ flux of a given
         cluster. In this method, we equally distribute the total SZ flux of
         equation (\ref{szflux}) among $N_{MC} = 4000$ random points, whose 3D
         density follow the ICM pressure, $P_g(r) \propto f(r/\rvir)\cdot g(r/\rvir)$, around the center of
         a given cluster. While the method would be equivalent to exact
         projection in the limit $N_{MC} \rightarrow \infty$, the
         poisson error introduced due to a finite value of $N_{MC}$ will be
         negligible comparing to the WMAP detector noise. The
         projected distribution of points should be then smeared by the WMAP beam window to get the
         expected distribution of the SZ flux. The expected SZ signal of
         the cluster in pixel $i$
         is proportional to $n_i$, the number of points that will fall into that
         pixel:
        \beq
         S_{i}(\nu)= \overline{\delta T}_{SZ, i}\cdot \delta \Omega_{\pix} =
         \left(\frac{n_i}{N_{MC}}\right) \times S_{tot}(\nu).
        \eeq
\subsection{Point Source Contamination}\label{secps}

         The frequency dependence of WMAP small angle anisotropies
         have been interpreted as a random distribution of point
         sources with a flat spectrum \citep[i.e. Antenna temperature scaling as $\nu^{-2}$;][]{WMAPfor}.
         The majority of individually identified WMAP point sources are also
          consistent with a flat spectrum.
          Since the SZ signal has a small frequency dependence at the range of WMAP frequencies ($41 ~{\rm GHz} <
         \nu < 94 ~{\rm GHz}$), we can use this frequency dependence
         to distinguish the Point Source (PS) contamination from the SZ
         signal. To do this, we assume a microwave point source with a flat (constant) luminosity per
         unit frequency, $L_{PS}$, for each cluster galaxy, and that the galaxies follow the dark
         matter distribution (equation \ref{NFW}) inside each
         cluster. The total microwave flux associated with the point
         sources is then given by
         \beq
            \overline{\delta T}_{PS} \cdot \delta \Omega \equiv \int_{\delta \Omega} \delta T_{PS} (\nu; {\bf \hat{n}})
        ~d^2{\bf \hat{n}} = \frac{2h^2 c^2}{k^3_B T^2_{\cmb}}\cdot \frac{\sinh^2(x/2)}{x^4}\left(\frac{N_{\rm vir} L_{PS}}{4\pi
        d^2_L}\right);~x= {h~\nu\over k_B T_{\cmb}},\label{ps}
          \eeq
        where $N_{\rm vir}$ is the number of galaxies, above a
% NOTE: add "luminosity distance"
        certain magnitude limit, within the virial radius, and $d_L$ is the
luminosity distance. For our analysis, we use the
% NOTE: add "K-band"
        \citet{lin04} result for 2MASS near infrared $K$-band selected galaxies:
        \beq
        N_{\rm vir}(M_K \leq -21) = 37 \pm 3 \left(\mvir \over 7
        \times 10^{13} h^{-1} \msun\right)^{0.85 \pm 0.04}.\label{eq:nm}
        \eeq
         Thus, $L_{PS}$ is defined as the total point
        source luminosity per unit frequency associated with the
% NOTE: minor change K -> $K$
        cluster, divided by the number of galaxies brighter than the near infrared $K$-band magnitude
        of $-21$, within the virial radius of the cluster.

%\subsection{Merging Clusters}\label{mergclus}
\section{Statistical Analysis Methodology}\label{CMBstat}

         For a low resolution CMB experiment such as WMAP, the main sources of uncertainty in the SZ
         signal are the primary CMB anisotropies, as well as the detector
         noise. Since both of these signals are well described by
         gaussian statistics, we can write down the $\chi^2$ which describes
         the likelihood of observing a given model of the cluster
         SZ+PS
         profile (see \sec \ref{SZprofile}):
         \beq
         \chi^2 = \sum_{i,j;a,b}
         \left[T_{ia}-S_i(\nu_a)/\delta\Omega_{\pix}\right] C^{-1}_{ia,jb}
         \left[T_{jb}-S_j(\nu_b)/\delta\Omega_{\pix}\right],\label{chi2}
         \eeq
         where $a$ and $b$ run over WMAP frequency bands (i.e. Q, V, or W), and $i$ and $j$ run over the
         WMAP pixels. Here, $T_{ia}$ and $S_i(\nu_a)$ are the observed temperature
         and expected SZ+PS flux in pixel $i$ and band $a$, while
         $C_{ia,jb}$ is the covariance matrix of pixel temperatures:
         \beq
         C_{ia,jb}= n^2_{ia}\delta_{ij}\delta_{ab}+\sum_{\ell} \left(\frac{2\ell+1}{4\pi}\right)
         ~|W_{\pix}(\ell)|^2 ~W_{\rm beam}(\ell;a)W_{\rm beam}(\ell;b) ~C_{\ell}
         P_{\ell}(\cos \theta_{ij}).
         \eeq
          Here, $n_{ia}$ is the pixel detector noise,
          $W_{\pix}$ and $W_{\rm beam}$ are the HEALPix pixel and WMAP
% NOTE: add theta_{ij} definition
          beam transfer functions \citep{WMAPbeam}, $C_{\ell}$'s and $P_{\ell}$'s are the primary
          CMB multipoles and Legendre
          polynomials respectively, and $\theta_{ij}$ is the angular seperation
between the pixels $i$ \& $j$. We use CMBfast code \citep{cmbfast} in
          order to generate the expected values of $C_{\ell}$'s for the
          WMAP concordance \lcdm cosmology \citep{WMAP}.

% NOTE: slight wording change
          Because WMAP detector noise
          only varies on large angular scales, $n_{ia}$
          can be assumed to be almost constant if we limit the
          analyses to the neighborhood of a cluster, yielding
          \bea
           C_{ia,jb}\simeq \sum_{\ell} \left(\frac{2\ell+1}{4\pi}\right)
         ~|W_{\pix}(\ell)|^2~K_{ab}(\ell)~P_{\ell}(\cos\theta_{ij}),\\
         K_{ab}(\ell) = W_{\rm beam}(\ell;a)W_{\rm beam}(\ell;b) ~C_{\ell}
         +n^2_{a}\delta_{ab}\delta\Omega_{\pix},
          \eea
          where we used
          \beq
            \delta\Omega_{\pix}\sum_{\ell} \left(\frac{2\ell+1}{4\pi}\right)
         ~|W_{\pix}(\ell)|^2 P_{\ell}(\cos\theta_{ij})= \delta_{ij}.
         \eeq
          Now, it is easy to check that, in the small angle limit,
          we have
          \beq
            C^{-1}_{ia,jb} \simeq \delta\Omega^2_{\pix}\sum_{\ell} \left(\frac{2\ell+1}{4\pi}\right)
         ~|W_{\pix}(\ell)|^2~K^{-1}_{ab}(\ell)~P_{\ell}(\cos\theta_{ij}),
         \eeq

         We can again use the Monte-Carlo method, described at the
         end of \sec \ref{SZprofile}, to evaluate ${\bf C^{-1} S}={\bf C^{-1}W}_{\rm beam}
         {\bf S_0}$, where ${\bf S_0}$ is the raw projected SZ profile. To do
         so, instead of smearing ${\bf S_0}$ by the detector beam
         window,${\bf W}_{\rm beam}$,
         we can smear ${\bf S_0}$ by $ {\bf C}^{-1}{\bf W}_{\rm beam}$, which is given by
         \beq
         D_{ia,jb} = C^{-1}_{ia,kb}W_{{\rm beam};k,j;b} \simeq \delta\Omega^2_{\pix}\sum_{\ell} \left(\frac{2\ell+1}{4\pi}\right)
         ~|W_{\pix}(\ell)|^2~K^{-1}_{ab}(\ell)~W_{\rm beam}(\ell;b)~P_{\ell}(\cos\theta_{ij}).
         \eeq

        Since the SZ signal is dominant at small angles, and at the
        same time we want to avoid the non-trivial impact of the CMB
        masks on the covariance matrix inversion, we cut off
        $D_{ia,jb}$ if the separation of pixels $i$ and $j$ is
        larger than $\theta_D = 3^{\circ}$. We do not expect this to
        impact our analysis significantly, as the CMB fluctuations
        are dominated by smaller angles. As $T_i = S_i/\delta \Omega_{\pix}$ still minimizes $\chi^2$, this truncation
        cannot introduce systematic errors in our SZ or PS
        signal estimates. However, it may cause an underestimate of the covariance errors.
         In \sec\ref{error}, we introduce a Monte-Carlo
         error-estimation method to alleviate this concern.

% NOTE: add a comma
        The $\chi^2$, given in equation (\ref{chi2}), is quadratic in
        $\lambda_1 = B(c)f_{\rm gas} h$ and $\lambda_2 = L_{{PS}}$,
        and can be re-written, up to a constant, as \beq
        \chi^2 = F^{\alpha \beta} \lambda_{\alpha}\lambda_{\beta}  -2 A^{\alpha}
        \lambda_{\alpha}, \label{chi_conc}
        \eeq
        where
        \bea
        F^{\alpha \beta} = \delta\Omega^{-2}_{\pix}\sum_{i,j;a,b}
        S^{\alpha}_i(\nu_a)C^{-1}_{ia,jb}S^{\beta}_j(\nu_b),\\
        {\rm and}~~~ A^{\alpha} = \delta\Omega^{-1}_{\pix}\sum_{i,j;a,b}
        S^{\alpha}_i(\nu_a)C^{-1}_{ia,jb}T_{jb}.
        \eea
        Note that $S_i(\nu_a)= \lambda_{\alpha}S^{\alpha}_i(\nu_a)$
        is the sum of the SZ + PS flux contributions per pixel (derived in \sec \ref{SZprofile} and \sec \ref{secps})
        for all the clusters in the sample.

        After evaluating the coefficients $F^{\alpha \beta}$ and
        $A^{\alpha}$ via the Monte-Carlo method described
        above, the $\chi^2$ in Eq. (\ref{chi_conc}) can be minimized
        analytically to obtain the best fit values for the gas
        fraction and point source luminosity. After this minimization, the resulting $\chi^2$ can be
        used to constrain the value of the concentration parameter
        $\c$.

         \subsection{Error Estimates}\label{error}

            While the covariance matrix obtained from the $\chi^2$ in equation (\ref{chi2}) gives a
         natural way to estimate the errors, our Monte-Carlo based
         approximation of the covariance matrix, as well as its truncation beyond
         $\theta_D = 3^{\circ}$, may reduce the accuracy of our error
         estimates.  Another source of error, which is not included in the
         covariance method, is the uncertainty in observed X-ray
         temperatures.

         In order to obtain more accurate error estimates, we
         use our primary CMB power spectrum (from CMBfast), combined with the WMAP
         noise and beam properties to generate 99 Monte-Carlo Realizations of WMAP CMB
         maps in its three highest frequency bands (Q, V, \& W).
         Neglecting the contamination of cluster signals by
         background points sources, these maps can then be used to estimate the error covariance
         matrix for our $f_{\rm gas} h$ and $L_{PS}$ estimators, within an accuracy
         of $\sqrt{2/99}\simeq 14\%$.

         To include the impact of $T_{\rm x}$ errors in our Monte-Carlo error estimates,
         we assume an asymmetric log-normal probability distribution for the true
        temperature, ${\cal P}(T_{\rm x})$, which is centered at
          the observed value, $T^{obs}_{\rm x}$, and its extent on each side is given by the
            the upper/lower error of the observed temperature, $\delta T^{u}_{\rm x}$/$\delta T^{l}_{\rm x}$,
            i.e.
            \bea
                {\cal P}(T_{\rm x}) dT_{\rm x} = \exp\left\{-\frac{[\ln(T_{\rm x}/T^{obs}_{\rm x})]^2}{2
                \sigma^2}
                \right\} \frac{d \ln
                T_{\rm x}}{\sqrt{2\pi\sigma^2}}, \nonumber\\ {\rm where}~~ \sigma=
                \left\{ \begin{array}{ll} \ln(1+\delta T^{u}_{\rm x}/T^{obs}_{\rm x}) &
                \mbox{if $T_{\rm x}>T^{obs}_{\rm x}$}, \\
                -\ln(1-\delta T^{l}_{\rm x}/T^{obs}_{\rm x}) & \mbox{if $T_{\rm x}<T^{obs}_{\rm x}$}. \end{array}
                \right.
            \eea
            Therefore, in each Monte-Carlo realization, the temperature of each cluster
            is also randomly drawn from the
            above distribution, which is then used to construct the SZ/PS template for that cluster
            (\sec \ref{SZprofile}).

\section{Results}\label{results}

   In this section, we use the framework developed in \sec
   \ref{CMBstat}
   to combine the WMAP temperature maps with our cluster
   catalog. It turns out that about $30\%$ of our clusters are within
   $3^{\circ}$, and about $8\%$ within $1^{\circ}$ of another cluster in our sample,
   implying possible correlations
   between the signals extracted from each cluster. However, in order
   to simplify the analysis and interpretation of our data, we
   ignore such possible correlations, and thus assume that the
   values of $f_{\rm gas}$ and $L_{PS}$, obtained for each cluster is
   almost independent of the values for the rest of the sample. As
   we argue below, there is no evidence that this approximation
   may have biased our error estimates of global averages significantly.

     One of our clusters (A426; Perseus cluster) shows an 18$\sigma$
     ($L_{PS} = (1.69 \pm 0.09) \times 10^{29}~ h^{-2} {\rm erg/s/Hz/galaxy}$)
     signature for frequency dependent PS signal. It turns
% NOTE: slight wording change
     out that the 5th brightest microwave source detected by the WMAP
     team (WMAP\#94; NGC 1275) happens to be the brightest galaxy of the
     cluster. As this point source overwhelms the SZ signal,
     we omit A426 from our analysis, which leaves us
% NOTE: remove "only"
     with a sample of 116 X-ray clusters.

\subsection{Global ICM gas fraction and Point Source
Luminosity}\label{global}

The most straightforward application of the statistical framework
introduced in \sec \ref{CMBstat} is to obtain a global best fit for
the gas fraction $f_{\rm gas}$ and galaxy microwave luminosity
$L_{PS}$ for a given value of concentration parameter $\c$. Table
\ref{global_tab} shows the results of our global fits for nominal
values of $\c=3$ and $\c=5$, within different temperature cuts,
which are also compared with the estimates from our compiled X-ray
observations. Note that the lower value of $\c$ is probably
appropriate for the high end of the cluster masses/temperatures,
while the higher value may correspond to less massive clusters. To
get the X-ray gas fraction, the gas mass estimated from X-ray
observations (\sec \ref{catalog}) is divided by the virial mass
expected from observed $T_{\rm x}$ (Eq. \ref{tx}) for each value of
$\c$.

\begin{table}[t]
\begin{center}
\caption{\label{global_tab} The global best fit values to the gas
fraction and point source parameters for different temperature cuts,
and assuming $\c=3$ (top) or $\c=5$ (bottom), compared with X-ray
estimates (see the text). $\Delta\chi^2$ shows the relative
significance of the best fit with respect to the no-cluster (null)
model.}

\begin{tabular}{|c|c ||c |c |c || c|}
\hline
 $\c= 3$  & \# &$f_{\rm gas} h$ (X-ray) & $f_{\rm gas} h$ (SZ)    & $L_{PS}(10^{28} h^{-2} {\rm erg/s/Hz/gal})$& $\Delta\chi^2$\\%& $\beta$ & $\lambda$ & $N$ \\
\hline

all clusters           &116& $0.0608 \pm 0.0004$ & $0.077 \pm 0.011$ & $ 0.19 \pm 0.08$ & $-53.1   $\\
$T_{\rm x} \ge 3 ~\kev$ & 78& $0.0743 \pm 0.0005$ & $0.073 \pm 0.012$ & $ 0.07 \pm 0.16$ & $-47.8   $\\
$T_{\rm x} \ge 5 ~\kev$ & 38& $0.0954 \pm 0.0007$ & $0.086 \pm 0.015$ & $ 0.07 \pm 0.44$ & $-55.6   $\\
$T_{\rm x} \ge 8 ~\kev$ &  8& $0.1151 \pm 0.0016$ & $0.083 \pm 0.020$ & $ 0.06 \pm 0.49$ & $-27.2   $\\

\hline
\end{tabular}

\end{center}
%\end{table}

%\begin{table}[t]
\begin{center}
%\caption{\label{global_tab5} Similar to Table \ref{global_tab3}, but
%for $c=5$.}

\begin{tabular}{|c|c ||c |c |c || c|}
\hline
 $\c=5$  & \# &$f_{\rm gas} h$ (X-ray) & $f_{\rm gas} h$ (SZ)    & $L_{PS}(10^{28} h^{-2} {\rm erg/s/Hz/gal})$& $\Delta\chi^2$\\%& $\beta$ & $\lambda$ & $N$ \\
\hline

all clusters           &116& $0.0662 \pm 0.0004$ & $0.084 \pm 0.011$ & $ 0.24 \pm 0.08$ & $-57.3   $\\
$T_{\rm x} \ge 3 ~\kev$ & 78& $0.0810 \pm 0.0005$ & $0.084 \pm 0.013$ & $ 0.21 \pm 0.21$ & $-51.2   $\\
$T_{\rm x} \ge 5 ~\kev$ & 38& $0.1040 \pm 0.0008$ & $0.101 \pm 0.017$ & $ 0.40 \pm 0.53$ & $-60.6   $\\
$T_{\rm x} \ge 8 ~\kev$ &  8& $0.1254 \pm 0.0018$ & $0.094 \pm 0.021$ & $ 0.33 \pm 0.59$ & $-29.8   $\\

\hline
\end{tabular}
\end{center}
\end{table}

While the overall significance of our model detections are in the
range of $5-8\sigma$, we see that the significance of our SZ
detection is $\sim 8 \sigma$ for the whole sample, and there is a
signature of point source contaminations at $\sim 3\sigma$ level,
although we should note that there is a significant correlation
($70-80\%$) between the SZ and PS signals.

While the SZ signal is mainly due to massive/hot clusters, most of
the PS signal comes from the low mass/temperature clusters (compare
1st and 2nd rows in each section of Table \ref{global_tab}).
Therefore,  for the PS signal, the higher concentration value of
$\c=5$ might be closer to reality, putting the average microwave
luminosity of cluster members at \beq \langle L_{PS}\rangle (41~{\rm
% NOTE: < -> \le
GHz}; M_K\le-21) = (2.4 \pm 0.8) \times 10^{27} h^{-2} {\rm
erg/s/Hz/galaxy}, \eeq

% NOTE: remove parenthesis
 Surprisingly, this number is very close to the diffuse WMAP Q-band
 luminosity of Milky Way and Andromeda galaxy \citep{ALS}, i.e. $\simeq 2\times 10^{27} {\rm erg/s/Hz}$. {\it Therefore, assuming that
a significant fraction of cluster
 members have a diffuse emission similar to Milky Way, our observation indicates that,
 on average, nuclear (AGN) activity cannot overwhelm the diffuse microwave emission
 from cluster galaxies.} Nevertheless, models of microwave emission
 from faint (radiatively inefficient) accretion flows cannot be
 ruled out (see \sec\ref{rss}).

As an independent way of testing the accuracy of our error
estimates, we can evaluate the $\chi^2$ for the residuals of our
global fits for the whole sample (first rows in Table
% NOTE: our -> the
\ref{global_tab}). For $\c=3$ and $\c=5$, the residual $\chi^2$ for
our global fits are $252$ and $268$, respectively, which are
somewhat larger than (but within $2\sigma$ of) the expected range
for $2\times 116$ degrees of freedom, i.e. $230 \pm 22$. While this
may indicate $\sim 7\%$ underestimate of errors, it may at least be
partly due to the $T_{\rm x}$ dependence of $f_{\rm gas}$, which we
discuss in the next section. Since correlation of errors among close
clusters may decrease this value, while systematic underestimate of
errors tends to increase the residual $\chi^2$, we conclude that,
unless these two effects accidentally cancel each other, we do not
see any significant evidence (i.e. $> 10\%$) for either of these
systematics. Repeating the exercise for the hotter sub-samples of
Table \ref{global_tab} yields a similar conclusion.

Finally, we note that the X-ray and SZ values for $f_{\rm gas} h$
% NOTE: add some words
are always consistent at the $2\sigma$ level (see further discussion below).

%However, our gas fractions are significantly ($\sim 2.5\sigma$)
%lower than the cosmic average baryon fraction for \lcdm concordance
%model \citep{Spergel2003} $\ob h/\om = 0.12 \pm 0.01$. If taken at
%the face value, this would imply that about $30-40\%$ of the baryons
%have been removed from the diffuse ICM.

\subsection{Dependence on the Cluster Temperature}\label{deptemp}

\begin{table}[t]
\begin{center}
\caption{\label{binned_tab} The global best fit values to the gas
fraction and point source parameters for different temperature bins,
assuming $\c=3$ (top) or $\c=5$ (bottom), compared with X-ray
estimates (see the text). $\Delta\chi^2$ shows the relative
significance of the best fit with respect to the no-cluster (null)
model. Note that the last bin only contains one cluster, i.e.
A2319.}

% NOTE: the columns are re-arranged; are you happy with this?

%\begin{tabular}{|c |c |c ||c |c || c |c |}
\begin{tabular}{|c |c |c ||c |c | c ||c |}
\hline
%$f_{\rm gas} h$ (X-ray) &$f_{\rm gas} h$ (SZ) & $L_{PS}(10^{28} h^{-2} {\rm erg/s/Hz})$& $T_{\rm x} (\kev)$ & $\overline{T}_{\rm x}(\kev)$ & \# &$\Delta\chi^2$ \\
$T_{\rm x} (\kev)$ & $\overline{T}_{\rm x}(\kev)$ & \# & $f_{\rm gas} h$ (X-ray) &$f_{\rm gas} h$ (SZ) & $L_{PS}(10^{28} h^{-2} {\rm erg/s/Hz})$ &$\Delta\chi^2$ \\
\hline
0-2& 1.1&20 & $ 0.0139 \pm 0.0056$ & $0.256 \pm 0.167$ & $ 0.22 \pm 0.10$ & -5.0  \\
2-4& 3.3&44& $ 0.0485 \pm 0.0004$ & $0.030 \pm 0.036$ & $ 0.13 \pm 0.20$ & -0.7  \\
4-6& 4.7&28& $ 0.0740 \pm 0.0010$ & $0.030 \pm 0.030$ & $-0.05 \pm 0.54$ & -1.9  \\
6-8& 6.5&16& $ 0.0914 \pm 0.0009$ & $0.100 \pm 0.031$ & $-0.10 \pm 1.31$ &-27.1  \\
8-10& 8.6& 7& $ 0.1126 \pm 0.0018$ & $0.076 \pm 0.025$ & $-0.05 \pm 0.51$ & -16.5  \\
10-12&11.0& 1 & $ 0.1271 \pm 0.0039$ & $0.133 \pm 0.053$ & $ 2.89 \pm 2.91$ & -11.8  \\

\hline
\end{tabular}

\begin{tabular}{|c |c |c ||c |c | c ||c |}
\hline
$T_{\rm x} (\kev)$ & $\overline{T}_{\rm x}(\kev)$ & \# & $f_{\rm gas} h$ (X-ray) &$f_{\rm gas} h$ (SZ) & $L_{PS}(10^{28} h^{-2} {\rm erg/s/Hz})$ &$\Delta\chi^2$ \\
\hline

0-2& 1.1&20 & $ 0.0152 \pm 0.0061$ & $0.299 \pm 0.169$ & $ 0.29 \pm 0.11$ & -6.3  \\
2-4& 3.3&44& $ 0.0528 \pm 0.0005$ & $0.014 \pm 0.041$ & $ 0.19 \pm 0.26$ & -0.6  \\
4-6& 4.7&28& $ 0.0806 \pm 0.0011$ & $0.037 \pm 0.035$ & $-0.12 \pm 0.72$ & -2.4  \\
6-8& 6.5&16& $ 0.0996 \pm 0.0010$ & $0.123 \pm 0.036$ & $ 0.33 \pm 1.62$ & -29.9  \\
8-10& 8.6& 7& $ 0.1227 \pm 0.0020$ & $0.085 \pm 0.028$ & $ 0.14 \pm 0.63$ & -16.7  \\
10-12&11.0& 1 & $ 0.1385 \pm 0.0043$ & $0.148 \pm 0.051$ & $ 3.98 \pm 3.24$ & -14.5  \\

\hline
\end{tabular}

%
%\begin{tabular}{|c |c |c ||c |c || c |c |}
%\hline
%$f_{\rm gas} h$ (X-ray) &$f_{\rm gas} h$ (SZ) & $L_{PS}(10^{28} h^{-2} {\rm erg/s/Hz})$& $T_{\rm x} (\kev)$ & $\overline{T}_{\rm x}(\kev)$& \# &$\Delta\chi^2$ \\
%\hline
%$ 0.0139 \pm 0.0056$ & $0.256 \pm 0.167$ & $ 0.22 \pm 0.10$ &  0-2& 1.1&20& -5.0  \\
%$ 0.0485 \pm 0.0004$ & $0.030 \pm 0.036$ & $ 0.13 \pm 0.20$ &  2-4& 3.3&44& -0.7  \\
%$ 0.0740 \pm 0.0010$ & $0.030 \pm 0.030$ & $-0.05 \pm 0.54$ &  4-6& 4.7&28& -1.9  \\
%$ 0.0914 \pm 0.0009$ & $0.100 \pm 0.031$ & $-0.10 \pm 1.31$ &  6-8& 6.5&16&-27.1  \\
%$ 0.1126 \pm 0.0018$ & $0.076 \pm 0.025$ & $-0.05 \pm 0.51$ &  8-10& 8.6& 7&-16.5  \\
%$ 0.1271 \pm 0.0039$ & $0.133 \pm 0.053$ & $ 2.89 \pm 2.91$ & 10-12&11.0& 1&-11.8  \\
%
%\hline
%\end{tabular}
%
%\begin{tabular}{|c |c |c ||c |c || c |c |}
%\hline
%$f_{\rm gas} h$ (X-ray) &$f_{\rm gas} h$ (SZ) & $L_{PS}(10^{28} h^{-2} {\rm erg/s/Hz})$& $T_{\rm x} (\kev)$ & $\overline{T}_{\rm x}(\kev)$& \# &$\Delta\chi^2$ \\
%\hline
%
%$ 0.0152 \pm 0.0061$ & $0.299 \pm 0.169$ & $ 0.29 \pm 0.11$ &  0-2& 1.1&20& -6.3  \\
%$ 0.0528 \pm 0.0005$ & $0.014 \pm 0.041$ & $ 0.19 \pm 0.26$ &  2-4& 3.3&44& -0.6  \\
%$ 0.0806 \pm 0.0011$ & $0.037 \pm 0.035$ & $-0.12 \pm 0.72$ &  4-6& 4.7&28& -2.4  \\
%$ 0.0996 \pm 0.0010$ & $0.123 \pm 0.036$ & $ 0.33 \pm 1.62$ &  6-8& 6.5&16&-29.9  \\
%$ 0.1227 \pm 0.0020$ & $0.085 \pm 0.028$ & $ 0.14 \pm 0.63$ &  8-10& 8.6& 7&-16.7  \\
%$ 0.1385 \pm 0.0043$ & $0.148 \pm 0.051$ & $ 3.98 \pm 3.24$ & 10-12&11.0& 1&-14.5  \\
%
%\hline
%\end{tabular}
%

\end{center}
\end{table}

\begin{figure}
\includegraphics[width=0.5\linewidth]{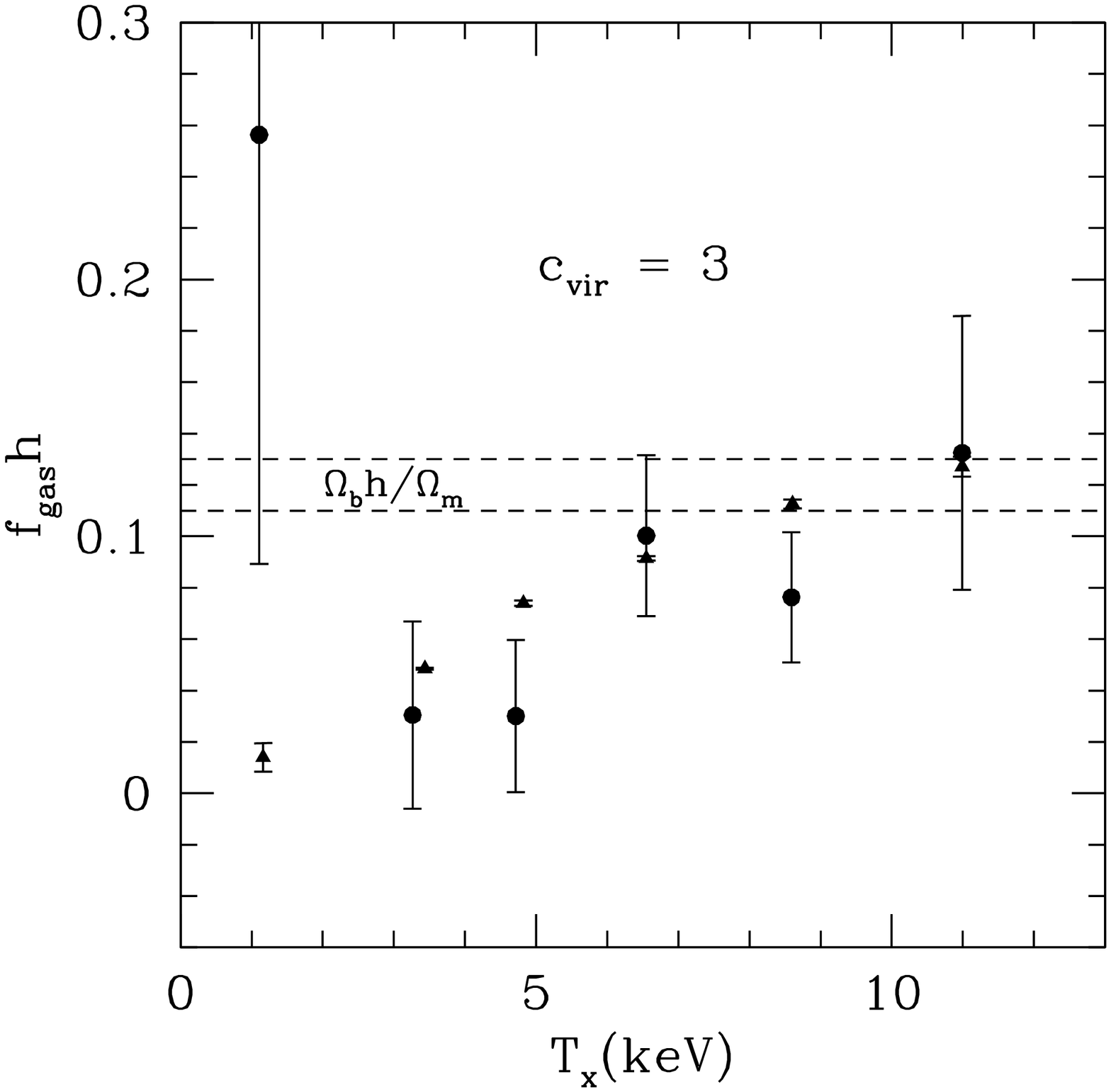}
\includegraphics[width=0.5\linewidth]{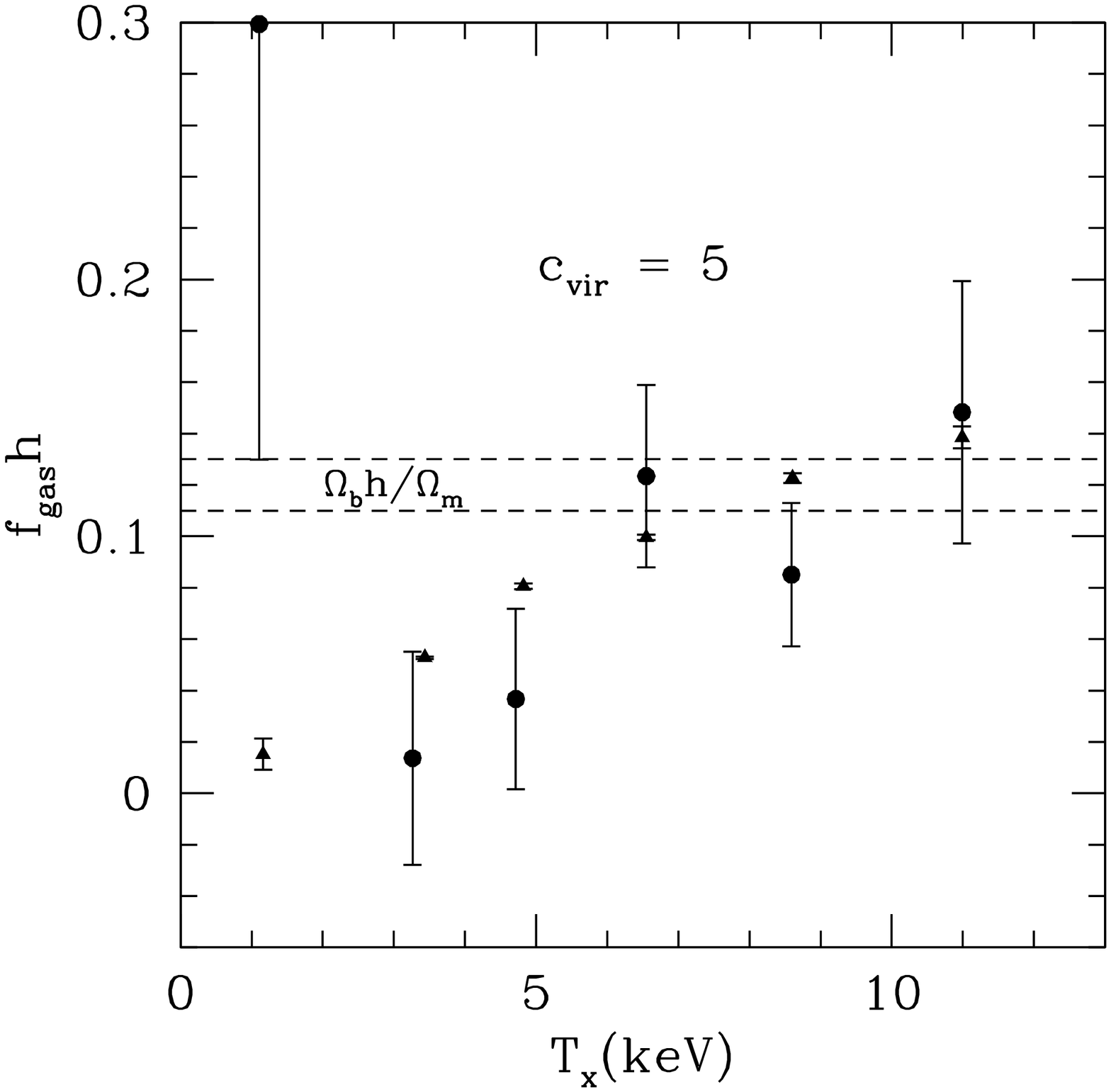}
\caption{\label{fgbin} The binned values of gas fractions (solid
circles/errorbars), inferred from our SZ measurements (assuming
$\c=3$ or $\c=5$; Table \ref{binned_tab}). The solid triangles are
the X-ray estimates for $f_{\rm gas}h$ (see the text), while the
region enclosed by the dashed lines is the upper limit from the WMAP
concordance model \citep{Spergel2003}.}
\end{figure}

\begin{figure}
\includegraphics[width=0.5\linewidth]{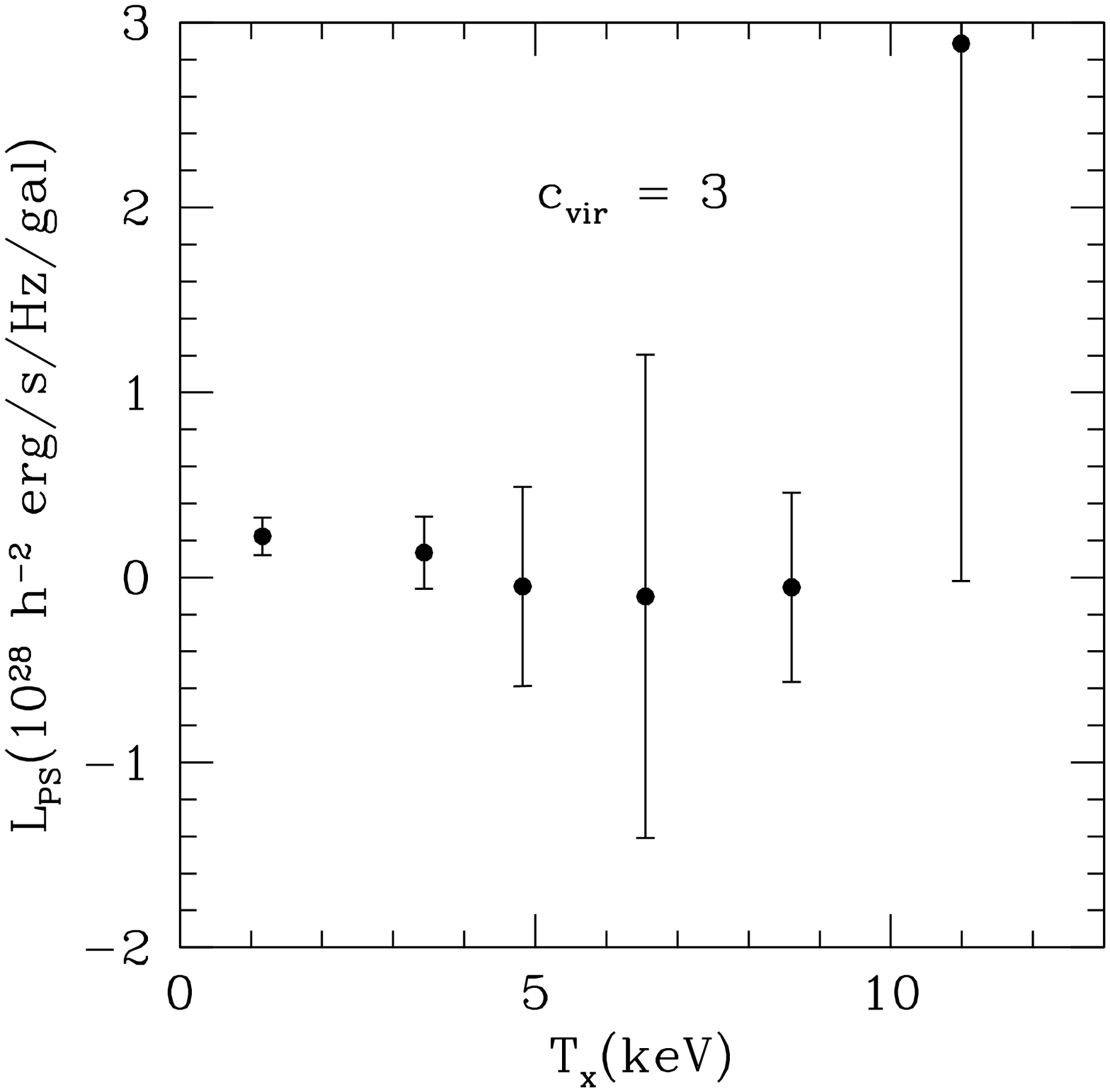}
\includegraphics[width=0.5\linewidth]{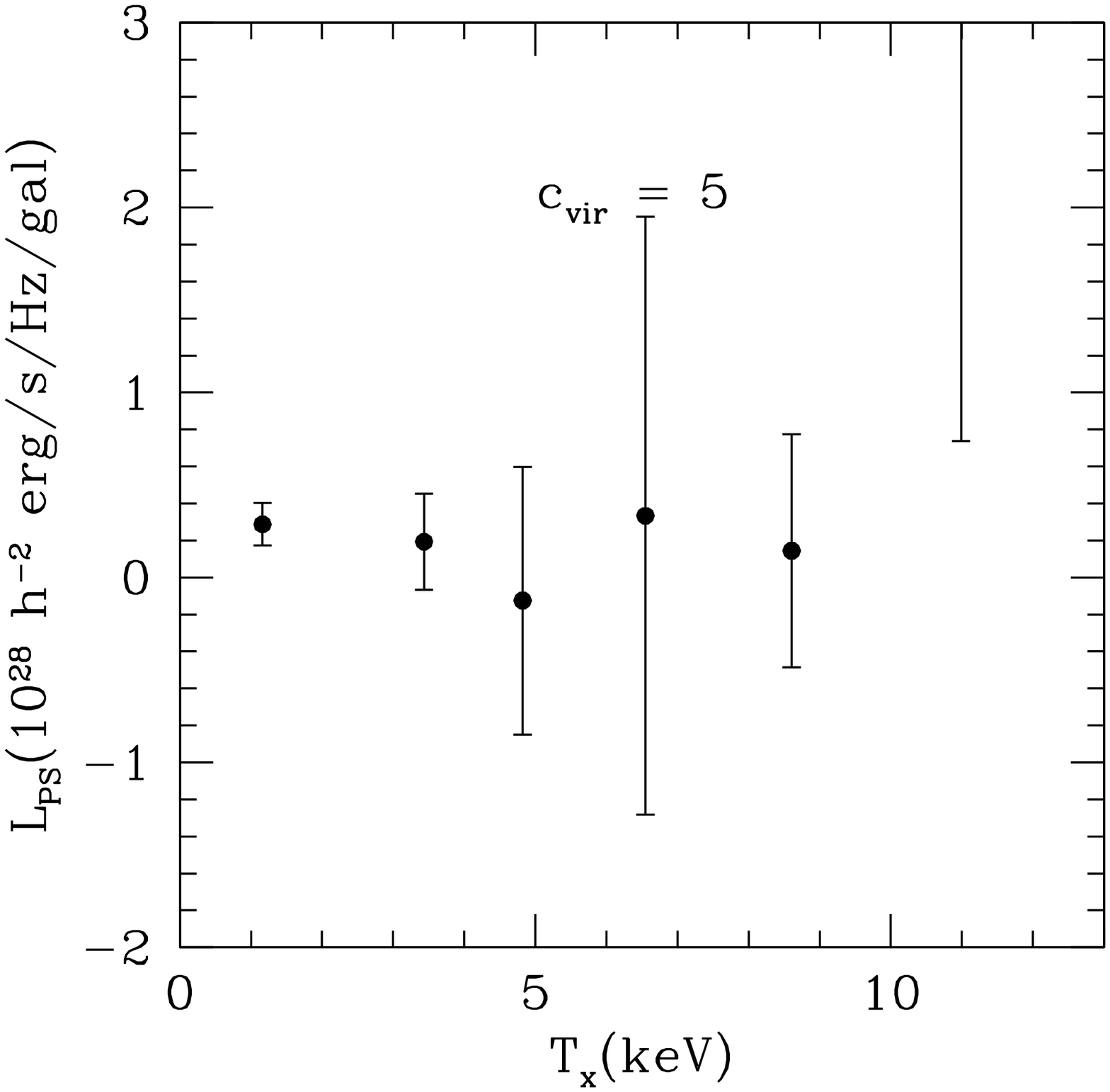}
\caption{\label{lrad} The binned values of the average microwave
luminosity per galaxy, $L_{PS}$, assuming a flat spectrum.}
\end{figure}

  Let us study the dependence of our inferred cluster
  properties on the cluster X-ray temperature, which can also be
  treated as a proxy for cluster mass (Eq. \ref{tx}). Since the
  errors for individual cluster properties are large, we average
  them within $2 ~\kev$ bins. The binned properties are
  shown in Figs.(\ref{fgbin}) \& (\ref{lrad}), and listed in Table
  \ref{binned_tab}. Similar to the previous section, we have also
  listed estimated gas fractions based on our compilation of X-ray
  observations.

    Fig. (\ref{fgbin}) compares our SZ and X-ray estimated gas
% NOTE: remove "errorbars"
    fractions. The solid circles show our SZ observations, while the triangles are the
    X-ray estimates.

     We notice that, similar to the global averages (Table
    \ref{global_tab}), our SZ signals are more or less consistent with the X-ray gas estimates.
    The $\chi^2$ for the difference of X-ray
    and SZ bins are $6.7$ and $7.6$ for $\c=3$ and $\c=5$ respectively,
    which are consistent with the 1-$\sigma$ expectation range of $6 \pm
    3.5$, for $6$ random variables. {\it Therefore, we conclude that
    we see no signature of any discrepancy between the SZ and X-ray estimates of
    the ICM gas fraction.}

    Another signature of consistency of our X-ray and SZ data points
    is the monotonically increasing behavior of gas fraction with
% NOTE: other -> previous
    $T_{\rm x}$\footnote{This trend is also responsible for the fact that
    a global fit (constant $f_{\rm gas}$) to the sample with $T_{\rm} > 3 ~\kev$
    is less significant than a global fit to the smaller sample with
    $T_{\rm x} > 5~ \kev$ (see Table \ref{global_tab}).}.

    Indeed, this behavior has been observed in previous
    X-ray studies \citep[e.g.,][]{mohr99,sanderson03}, and has been interpreted as a signature of preheating
    \citep{bialek2001} or varying star formation efficiency
    \citep{bryan2000}. A power law fit to our binned data points
    yields
    \bea
    f_{\rm gas}h = (0.069 \pm 0.014) \left(T_{\rm x} \over 6.6
    ~\kev\right)^{1.0^{+0.8}_{-0.6}} ~({\rm for}~\c=3), \label{fgast3}\\ ~ {\rm and}~ f_{\rm gas}h = (0.077 \pm 0.014) \left(T_{\rm x} \over 6.6
    ~\kev\right)^{1.1^{+0.8}_{-0.6}} ~({\rm for}~\c=5),\label{fgast5}
    \eea
    where the uncertainties in the normalization and power are almost
    un-correlated .

    Fig.(\ref{lrad}) shows that, after removing A426 ($T_x = 6.4$ keV), none of our bins show more than
    2-$\sigma$ signature for point sources. The fact that the
    observed amplitude of point sources changes sign, and is consistent
    with zero implies that any potential systematic bias of the SZ signal due to
    our modeling of the point sources (\sec \ref{secps}) must be
    negligible.

 %   values of LMS03 \beq
 %   f_{\rm gas} h = (0.068 \pm 0.002)~ T_{\rm x}(\kev)^{0.26\pm0.07}
 %   ~~{\rm (X-ray)}
%\eeq
 %    which is also consistent with the WMAP concordance upper limit \citep[$=\ob
  %  h/\om= 0.12 \pm 0.1$;][]{Spergel2003} for $T_{\rm x} = 7-10 ~\kev$.
  %  Fitting a similar power-law to our data results in
  %  \beq
  %  f_{\rm gas} h = (0.040 \pm 0.006)~ T_{\rm x}(\kev)^{0.26} ~~{\rm
  %  (SZ)},
  %  \eeq
  %  showing a $4.4\sigma$ discrepancy.

\subsection{SZ flux-Temperature relation}\label{szft}

    Given a perfect CMB experiment, and in the absence of primary anisotropies and foregrounds, in principle,
%NOTE: add "be"
    the SZ flux is the only cluster property that can be robustly measured from the CMB
    maps, and does not require any modeling of the ICM, while any
    measurement of the gas fraction would inevitably rely on the
    cluster scaling relations and/or the assumption of a relaxed
    spherical cluster. WMAP is of course far from such a perfect CMB
    experiment. Nevertheless, we still expect the SZ flux
    measurements to be less sensitive to the assumed ICM model (\sec \ref{model}),
    compared to our inferred gas fractions. Therefore, here we also
    provide a SZ flux-$T_{\rm x}$ scaling relation which should be
    more appropriate for direct comparison with other SZ
    observations and hydro-simulations. Plugging Eqs. (\ref{fgast3})
    and (\ref{fgast5}) into Eq. (\ref{szflux}) yields
% NOTE: eqnarray -> equation, are you okay with this?
    \beq
    S_{\rm vir}({\rm Jy})~ d^2_A(h^{-1} \mpc)~ E(z) = - (2.41 \pm
    0.49) \times 10^3 ~
    L(x)
    \left(T_{\rm x} \over 6.6
    ~\kev\right)^{3.5^{+0.8}_{-0.6}} ~({\rm for}~\c=3), \label{szt3}
    \eeq
 and
    \beq
    S_{\rm vir}({\rm Jy})~ d^2_A(h^{-1} \mpc) ~E(z)= - (2.04 \pm 0.37) \times 10^3 ~L(x) \left(T_{\rm x} \over 6.6
    ~\kev\right)^{3.6^{+0.8}_{-0.6}} ~({\rm for}~\c=5),\label{szt5}
    \eeq
    where
    \beq
    L(x)= \frac{x^4 (x\coth(x/2)-4)}{\sinh^2(x/2)},
    \eeq
    and $x=h\nu / k_B T_{\cmb}$ is the detector frequency in units
    of the CMB temperature.
% NOTE: wording change
  For the three highest
    frequencies of WMAP,  Q(41 GHz), V(61 GHz), and
  W(94 GHz), $L(x)=3.8,7.6,$ and $13.7$ respectively.
   The fits should hold within $3~\kev \lesssim T_{\rm x} \lesssim 11~\kev$, which
   is the range of cluster temperatures which contribute the most to
   our SZ detection. Notice that the difference between the
   normalizations inferred for two concentrations is comparable to
   the measurement errors.

     \citet{benson} is the only other group which expresses its SZ observations in terms of
     SZ flux-temperature relation. Our result is consistent with their observations, within
     the relevant temperature range ($T_{\rm x} \sim 9 ~\kev$
      for their sample). This is despite the higher median redshift
      of their sample ($z \sim 0.2-0.8$), which may indicate no
      detectable evolution in the SZ flux-temperature normalization. Their
      scaling with temperature, however, is significantly shallower
      than our measurement ($\propto T^{2.2\pm 0.4}_{\rm x}$), which is in
      contrast with our scaling ($\propto T^{3.5}_{\rm x}$), at more than $2\sigma$
      level. This is most likely due to the difference in the range of
% NOTE: wording change
      temperatures that are covered in our analysis. Indeed, the clusters in our
      three highest temperature bins whose temperature coincides with that
      covered in \citet{benson} show a much shallower
      dependence on temperature (see Fig. \ref{fgbin}), which is consistent with their
      results.

      We should note that we have to use our ICM model of \sec \ref{model} to
% NOTE: add some words
      convert $S_{\rm vir} = S_{200}$ to a flux within a much smaller area, $S_{2500}$, which is
      reported in \citet{benson}. The conversion factor is $0.30$
      and $0.42$ for $\c=3$ and $5$ respectively.

      As to comparison with numerical simulations of the ICM, even
      the most recent studies of the SZ effect in galaxy clusters
      \citep{dasilva,diaferio}
      include only a handful of clusters above $T_{\rm x} = 5~\kev$. This is
      despite the fact that most observational studies of the SZ
      effect, including the present work, are dominated by clusters
      with $T_{\rm x} > 5 ~\kev$. Therefore, a direct comparison of
      our observed SZ fluxes with numerical studies is not yet
      feasible. Instead, we can compare the SZ fluxes for clusters
      around $T_{\rm x} = 5 ~\kev$, where the temperature range of observed and
      simulated clusters overlap. Making this comparison, we see that,
% NOTE: wording change, are you ok with this?
      for the few simulated clusters with $T_{\rm x} \sim 5~\kev$ in \citet{dasilva} the SZ fluxes
      are in complete agreement with our observations. However, clusters of
      \citet{diaferio} are underluminous in SZ by close to an order of
      magnitude. Indeed, \citet{diaferio} also notice a similar
      discrepancy with the SZ observations of \citet{benson}. The
      fact that the results of \citet{diaferio} are inconsistent with other simulations and
      observations, may be indicator of a flaw in their analysis.

\subsection{Constraining the Concentration Parameter}

    It is clear that the assumption of constant concentration parameter, $\c$, which we have adopted
    up to this point,
    is an oversimplification. The average value of the concentration
    parameter is known to be a weak function of the cluster mass in CDM simulations
    \citep[$\propto M^{-0.1}$; e.g., NFW,][]{Eke2001}; even for a given
    mass, it follows a log-normal distribution \citep{bullock,dolag}, which
    may also depend on mass \citep{afshordi2002}.

\begin{figure}
\includegraphics[width=\linewidth]{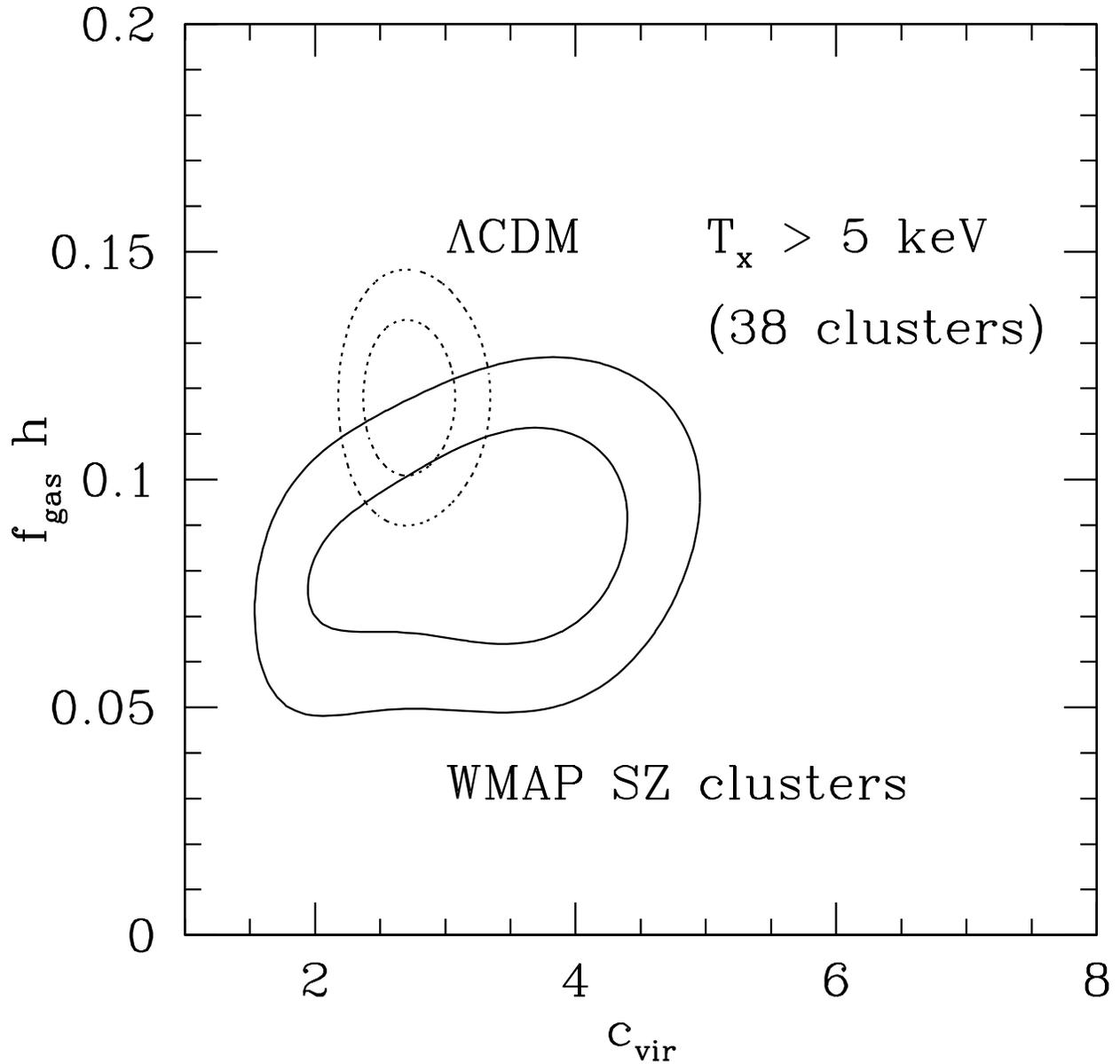}
\caption{\label{cfh} The $68\%$ and $95\%$ likelihood contours in
the $f_{\rm gas} h-c_{\rm vir}$ plane for clusters with $T_{\rm x} >
5~\kev$. The solid contours show the result from our SZ analysis, as
described in the text. The dotted contours are for the \lcdm
concordance model, where $f_{\rm gas} h$ is from
\citet{Spergel2003}, while the concentration parameter and its
uncertainty are obtained through the formalism developed in
\citet{afshordi2002}.}
\end{figure}

    As discussed at the end of \sec \ref{CMBstat}, we can repeat our
    Monte-Carlo template making procedure for different values of
    $\c$, which yields quadratic expressions for $\chi^2$, and thus
    enables us to (after marginalizing over $L_{PS}$) draw likelihood contours in the $f_{\rm gas}h-\c$ plane.
    From Table (\ref{global_tab}), we see that most of our SZ signal is due to the clusters with $T_{\rm x}
    \ge 5~\kev$ (see also Table \ref{binned_tab}), while $\c$($\propto T_{\rm x}^{-0.15}$) is expected to stay reasonably constant in this
    range. Therefore, we restrict the analysis to this sample.
    Fig. (\ref{cfh}) shows the result, where we have explicitly computed the
    $\chi^2$ for integer values of $\c$, and then interpolated it for
    the values in-between. The solid contours show our $68\%$ and $95\%$
    likelihood regions ($\Delta \chi^2 =2.2$ and $6.2$). We see that our data
    can constrain the concentration parameter of the dark matter
    halos to $c_{\rm vir}=3.4^{+0.6}_{-0.9}$ (median $\pm$ $~68\%$ percent likelihood).
    The dotted contours show the same likelihoods expected for clusters hotter than $5~\kev$
    in the WMAP \lcdm concordance model, where $f_{\rm gas}h = \ob h/\om = 0.12 \pm
    0.01$ is the upper limit expected from the WMAP concordance cosmology \citep{Spergel2003},
    while the range of $c_{\rm vir}$ is based on an extension of the
    top-hat model \citep{afshordi2002} for the same cosmology, averaged over the masses\footnote{The relation between
    cluster masses and temperatures (Eq. \ref{tx}) is a function of $c_{\rm vir}$ itself, but the uncertainty introduced
    in cluster mass estimates as a result, only slightly affects the obtained concentration range.}
    of our cluster sample, and inversely weighted by the square of
    temperature errors.

     We see that, while the mean gas fraction is about $30\%$ ($\sim
     2\sigma$) smaller than the cosmic upper limit (see the 2nd row in Table \ref{global_tab}),
     our inferred constraint
     on $c_{\rm vir}$ is completely consistent with the \lcdm
     prediction.

\section{Discussions}\label{discuss}

\subsection{ICM Gas Fraction}
In \sec \ref{results}, we demonstrated that, while we are able to
detect the thermal SZ effect at the 7-8$\sigma$ level from the first
year data release of WMAP temperature maps, the inferred gas
fractions are typically smaller than the X-ray estimates, as well as
the cosmological upper limit ($=\ob h/\om= 0.12 \pm 0.01$). The SZ
(as well as X-ray) observations have been often used, in combination
with the nucleosynthesis bound on $\ob$, to constrain $\om$, through
replacing the upper limit by equality
\citep{Myers1997,Mason2001,Grego2001,lancaster}. Nevertheless,
similar to our finding, such determinations have consistently
yielded lower values than, the now well-established, upper limit. In
fact, cooling and galaxy formation do lead to a depletion of the ICM
gas. To make the matters worse, supernovae feedback can make the gas
profile shallower, also leading to smaller baryonic fraction within
a given radius. {\it After all, clusters may not be such accurate
indicators of the baryonic census in the universe.}
% NOTE: I've added some sentences here; are you ok with this?

We note that these processes will affect low mass clusters more
strongly than high mass ones, and therefore it is natural to expect
the massive clusters to be better representative of the cosmic
baryonic content.
%As shown by \citet{lin03}, when the stellar mass is included, baryon
%mass fraction in massive clusters (e.g. $T_x \ge 4$ keV) is in good
%agreement with the WMAP determination.
% NOTE: X-ray only -> X-ray-based
Indeed, X-ray-based determinations of the ratio of cluster gas to
virial mass (for massive clusters) have seemingly been more
successful in reproducing the cosmic average \citep[after $\sim
10\%$ correction for stars; e.g.,][]{lin03}. However, we should note
that, as the X-ray emissivity is proportional to the square of local
plasma density, any smooth modeling of the ICM which may be used to
infer the gas mass from the X-ray map of a cluster, tends to
overestimate this value due to the contribution of unresolved
structure to the X-ray emissivity (i.e. $\langle n^2 \rangle >
\langle n \rangle ^2$). The hydrodynamical simulations can suffer
from the same problem, and thus fail to estimate the full magnitude
of the effect.
%We believe this can be the reason for larger values of $f_{\rm gas}$
%in X-ray observations, while the real ICM gas fractions may well be
%below the cosmic average.

There are several factors which can account for the discrepancy
between the gas fractions determined here and those estimated from
previous X-ray studies. Firstly, since direct detection of X-rays
from the ICM is rarely possible near the virial radius, a
significant degree of extrapolation is required to infer gas
properties at $r_{200}$. X-ray studies typically assume a
$\beta$-model form for the gas distribution, with an
empirically-motivated index parameter of $\beta\sim 2/3$
\citep[e.g.,][]{jones99}, implying $\rho_{{\rm gas}} \propto r^{-2}$
at large radii. By contrast, our physically-motivated model for the
gas density (Eq.\ref{rhogas}) yields $\rho_{{\rm gas}} \propto
r^{-3}$ at large radii, which produces a lower gas mass within
$r_{200}$. Secondly, the dark matter concentration is likely to be
higher than both values assumed here in less massive halos, as a
consequence of hierarchical formation. This underestimation of
$c_{\mathrm{vir}}$ correspondingly underestimates
$f_{\mathrm{gas}}$. Thirdly, the effects of non-gravitational
heating on the ICM in cooler clusters can act to displace the gas
beyond the radius where we observe it directly
\citep{mohr99,sanderson03}. Consequently, extrapolating to
$r_{200}/r_{500}$ based on the resulting lower-density gas that is
observed will lead to an underestimate of the total gas mass.

%Thirdly, the effects of non-gravitational heating on the gas in
%cooler clusters may well push the accretion shock beyond $r_{200}$,
%thus decreasing the total gas mass within this radius
%\citep{mohr99,sanderson03}.

The difference in logarithmic slope of $\rho_{\mathrm{gas}}(r)$ at
large radius between a $\beta$-model with $\beta\sim 2/3$ and Eq.
(\ref{rhogas}) is partly due to the effects of non-gravitational
physics biasing the gas distribution with respect to the dark
matter. However, there is some evidence that $\beta$ may increase at
larger radius in some \citep{vikhlinin99}, although not all
\citep{sanderson03} clusters, which would reduce the discrepancy.
However, satisfactory resolution of this issue will require mapping
of the gas distribution out to the virial radius in a representative
sample of clusters; a task which is complicated by the large angular
size and low surface brightness of the outer regions of the ICM.

%\subsection{Concentration Parameter}
\subsection{Model Uncertainties}\label{model_unc}

The relationship between the observed SZ flux and the ICM gas
fraction (Eq. \ref{szflux}), relies on the accuracy of the
(electron) virial temperature-mass relation (Eq. \ref{tvir}).
Although our X-ray temperature-mass relation (Eq. \ref{tx}) is
consistent with observations, the (emission weighted) X-ray
temperature only probes the inner parts of the cluster, and there
can still be significant deviations from our simple picture of the
uniform ICM in the cluster outskirts.

For example, \citet{voit} argue that, compared to a uniform
homogeneous accretion, inhomogeneous accretion will inevitably lead
to smaller entropy production. Although this may not significantly
affect the central part of a cluster, it can significantly change
the boundary condition (see Eq. \ref{temp}) behind the accretion
shock. This can be also interpreted as incomplete virialization
which can lead to smaller virial temperatures.

Let us estimate how much error the model uncertainties may introduce
to our measurements. Neglecting the contribution of radio sources
(see \sec\ref{rss}), we can divide the model uncertainties into the
SZ profile shape, and SZ flux uncertainties.

 In \sec\ref{results}, we saw that assuming $\c=5$ instead of
$\c=3$ may result in $\sim 10\%$ difference in the inferred gas
fraction. With the exception of merging clusters, given the low
resolution of WMAP maps, this is the level of error that we expect
may be introduced due to the uncertainty in the profile shape.

 As to the SZ flux uncertainty in our model (Equation
 \ref{szflux}), we note that the inferred $T_{\rm x}$ dependence of
 the SZ flux hinges upon the accuracy of our X-ray temperature-mass
 relation (Equation \ref{tx}). Parameterizing this as
 \beq \mvir = A T^{3/2}_{\rm x}, \eeq in Appendix A, we argue that
 the systematic uncertainty/error in $A$ is $\sim 10\%$, i.e. $\Delta A/A \sim 0.1$.
As the total SZ flux is proportional to the estimated virial mass,
we also get \beq {\Delta S_{tot}\over S_{tot}} = {\Delta A\over A}
\sim 0.1.\eeq Moreover, the virial radius of the cluster, which is
crucial in our matched filter method, is modified according to \beq
\rvir \propto \mvir^{1/3} = A^{1/3} T^{1/2}_{\rm x}, \eeq yielding
\beq \frac{\Delta \rvir}{\rvir} = \frac{\Delta\c}{\c} = \frac{1}{3}
\left(\Delta A\over A\right) \sim 0.03. \label{dcv}\eeq The SZ
template is then given by \beq S(\theta) = \left(S_{tot} d^2_{A}
\over \pi \rvir^2\right) \Sigma\left(\theta d_{A}/ \rvir\right),
\eeq where $\Sigma(x)$ is the normalized SZ (integrated pressure)
profile in our model\footnote{Note that the gas pressure is
proportional to $f(x)g(x)$; see Appendix A.}: \beq \Sigma(x) =
{\int^{1}_{x} f(y)g(y)\frac{y dy}{\sqrt{y^2-x^2}}\over 2 \int^1_0
f(y)g(y)y^2dy}. \eeq Therefore, the systematic error in the
estimated $f_{\rm gas}$ is given by: \beq \frac{\Delta f_{\rm
gas}}{f_{\rm gas}} = -{\Delta S \over S} =
\left[-1+\frac{1}{3}\left(2+{d\ln \Sigma\over d\ln x}\right)\right]
{\Delta A\over A} \simeq -{\Delta A \over A} \sim -0.1, \eeq where
we used Equation (\ref{dcv}), and the fact that $d\ln \Sigma /d\ln x
\simeq -2$ in cluster outskirts, where most of the SZ signal comes
from. Therefore, we see that the expected level of systematic error
due to the theoretical uncertainty in the SZ profile/flux is $\sim
15\%$, which is comparable to our random error. However, the similar
systematic error in our constraints on $\c$ ($\sim 3\%$; Equation
\ref{dcv}), is significantly smaller than the associated random
error.

 Further complication may be introduced by
 the fact that, as a result of ICM inhomogeneities and incomplete
 frequency coverage, the observed X-ray temperature, $T_{\rm x}$, could be different from
 the emission weighted temperature, $T_{ew}$, derived in Appendix A.
 Recently, \citet{Mazzotta} defined a {\it spectral-like} temperature, $T_{sl}$,
 which can approximate the observed $T_{\rm x}$ to better than a few
 percent, and \citet{Mazzotta} claim that, in hydrodynamic N-body simulations,
 $T_{sl}$ can be lower than $T_{ew}$ by as much as 20-30\%.
 However, applying the definition of $T_{sl}$ to the
 analytic model of Appendix A (which is clearly less structured than both simulations and observations),
 we find that $T_{sl}$ is smaller than $T_{ew}$ by only $\sim 5\%$.
 Therefore, we can ignore the impact of this discrepancy in our
 analyses.

%we expect model uncertainties, such as asphericity, incomplete
%virialization, or deviations from the polytropic relation to affect
%our results at a similar level.

Finally, another possibility is the breakdown of Local Thermal
Equilibrium (LTE) in the cluster outskirts. While the hydrodynamic
shocks heat up the ions instantly, the characteristic time for
heating up the electrons (via Coulomb interactions) can be
significantly longer \citep{foxloeb,chieze,takizawa}, and thus the
electron temperature can be lower by as much as $20\%$ in the outer
parts of clusters. Since the thermal SZ effect is proportional to
the electron temperature, the breakdown of LTE can be a source of
low SZ signals. Interestingly, the effect is expected to be bigger
for more massive clusters which have longer Coulomb interaction
times.

\subsection{Radio Source spectrum}\label{rss}

While we assumed an exactly flat spectrum for all our point source
contamination, a more realistic model would include a random spread
in the spectral indices $\alpha$ of the point sources. In fact,
although the average spectral index of the WMAP identified sources
is zero \citep[flat;][]{WMAPfor}, there is a spread of $\sim 0.3$ in
$\alpha$ of individual sources. Moreover, although most of the
bright microwave sources have an almost flat spectrum, an abundant
population of faint unresolved sources may have a different spectral
index, and yet make a significant contribution to the cluster
microwave signal. For example, a VLA survey finds that for sources
with flux $S\ge 0.1$ Jy, the mean spectral index distribution
(within $8\le \nu \le 90$ GHz) can be described by a Gaussian whose
mean and dispersion are $-0.37$ and 0.34, respectively
\citep{holdaway94}. At fainter flux limit ($\sim 20$ mJy), the CBI
experiment finds that the mean index (from 1.4 to 31 GHz) is
$-0.45$, with maximum and minimum indexes being 0.5 and $-1.32$,
respectively \citep{mason03}.

In order to test the sensitivity of our SZ signal to the point
source spectrum, we repeat the analysis for the spectral power
indices of $-1$ and $1$. Our over-all SZ signal changes by less than
$2\%$, implying the insensitivity of our results to the assumed
spectrum.

The average luminosity in the microwave band of cluster galaxies
also does not sensitively depend on our choice of the spectral
index. For example, assuming $\alpha=-0.8$, as most studies in low
frequencies adopt \citep[e.g.,][]{cooray98}, we find that at 41 GHz
the luminosity changes less than 10\% to $\langle L_{PS}\rangle =
(2.2 \pm 0.8) \times 10^{27} h^{-2} {\rm erg/s/Hz/galaxy}$.

Finally, it is interesting to compare our inferred galaxy luminosity
at 41 GHz with that observed at lower frequencies. A recent study of
a large sample of nearby clusters has determined the cluster AGN
luminosity function (LF) at 1.4 GHz \citep{lin05}, which is in
agreement with the bivariate LF obtained by \citet{ledlow96}; it is
found that the cluster LF is very similar to that of the field, once
the difference in the overdensity has been taken into account.
Integrating the LF from $10^{27}$ to $10^{33.5}$ erg/s/Hz
(corresponding to the observed luminosity range of AGNs) and
multiplying by the cluster volume gives the total luminosity
$L_{1.4}$ at 1.4 GHz. For a $5\times 10^{14} M_\odot$ cluster,
$L_{1.4} = 2.43\times 10^{31}$ erg/s/Hz, which can be compared to
the total luminosity inferred from our result $L_{tot}=\langle
L_{PS}\rangle N_{vir} (1.4/41)^\alpha$ (c.f. Eqn~\ref{eq:nm}).
Assuming $\alpha=0$, we find $L_{tot}=7.12\times10^{29}$ erg/s/Hz;
with $\alpha=-0.8$, $L_{tot} = 9.72\times 10^{30}$ erg/s/Hz.

This exercise suggests that a typical spectrum of $\alpha(1.4,41)
\sim -1$ brings our results into good agreement with the 1.4 GHz
measurements. However, this does not necessarily imply the spectral
shape to be similar within the frequency range that is relevant to
our analysis (41 to 94 GHz). The fact that changing $\alpha(41,94)$
from 0 to $-0.8$ affects $\langle L_{PS}\rangle$ less than 10\%
suggests our analysis is robust against specific choices of the
spectral index.

An alternative to this picture is drawn in \citet{pierpaoli}, where
it is proposed that a significant part of the point source microwave
emission in sky may come from faint (radiatively inefficient)
accretion flows around supermassive black holes in early-type
galaxies. The assumed microwave luminosity per early-type galaxy in
their model is $\sim 10^{28} {\rm erg/s/Hz}$, which is consistent
with our results, if less than $50\%$ of cluster galaxies with $M_K
\le -21$ are early-type galaxies with microwave luminosities at this
level. Using the luminosity function of near infrared galaxies, we
find that the number density of early-type galaxies \citep{kochanek}
with $M_K \le -21$ ($\simeq 7.6 \times 10^{-3} h^3 \Mpc^{-3}$) is
indeed close to the number density of early-types assumed in
\citet{pierpaoli}, which is $5.8 \times 10^{-3} h^3 \Mpc^{-3}$. A
nearby optical study of galaxy population in clusters suggests that
the early-type fraction in clusters is $\sim 20\%$, and does not
show strong variation with cluster-centric distance \citep{goto}.
Based on this finding, we compute the average early-type fraction
for galaxies with $M_K \leq -21$ from the type-specified luminosity
functions of \citet{kochanek}, and assume that it roughly stays the
same within cluster virial radii. The obtained fraction is $\simeq
34\% (<50\%)$, which suggests that our observed microwave point
source luminosity is consistent with the characteristics of the
faint accretion flows proposed in \citet{pierpaoli}.
%We then find that the average fraction of early-type galaxies with
%$M_K \le -21$ is $\simeq 34\%$ \citep{kochanek}, and does not change
%significantly within cluster virial radii \citep{goto}. Therefore,
%we conclude that our observed microwave point source luminosity is
%consistent with the faint accretion flows proposed in
%\citet{pierpaoli}

\section{Conclusions and Future Prospects}\label{conclusions}

  In this work, using
  a semi-analytic model of the Intra-Cluster Medium,
  we devised a Monte-Carlo based optimal filter match method to extract the thermal
  SZ signal of identified X-ray clusters with measured X-ray temperatures.
  We apply the method to a catalog of 116 low redshift X-ray
  clusters, compiled from the literature, and detect the SZ signal, at
  $\sim 8\sigma$ level (random error), while we estimate a comparable systematic error due to model uncertainties.
   We also see a $3\sigma$ signature for
  point source contamination, which we model based on the assumed
  spectral characteristics and spatial distribution of point
  sources. It turns out that the average luminosity of bright cluster
% NOTE: < -> \le
  members ($M_K \le -21$) is comparable to that of Milky Way and Andromeda.
  While our observed SZ signal constrains the gas fraction of the
   Intra-Cluster Medium to $60-70\%$ of the cosmic average, it is completely
   consistent with the gas fractions based on our compiled X-ray gas mass estimates.

   Based on our results, we also derive the SZ flux-temperature
   relation within the temperature range of $4~\kev < T_{\rm x} < 11~\kev$, and
   compare it with other numerical/observational studies. While our
   findings are consistent with other SZ observations, the range of
   cluster temperatures covered by numerical simulations is too low
   to permit any meaningful comparison.

    Finally, after marginalizing over gas fraction and point source
    contaminations, we could constrain the average dark matter halo concentration
    parameter of clusters with $T_{\rm x} > 5~\kev$ to $c_{\rm
    vir}=3.4^{+0.6}_{-0.9}$.

\begin{figure}
\includegraphics[width=\linewidth]{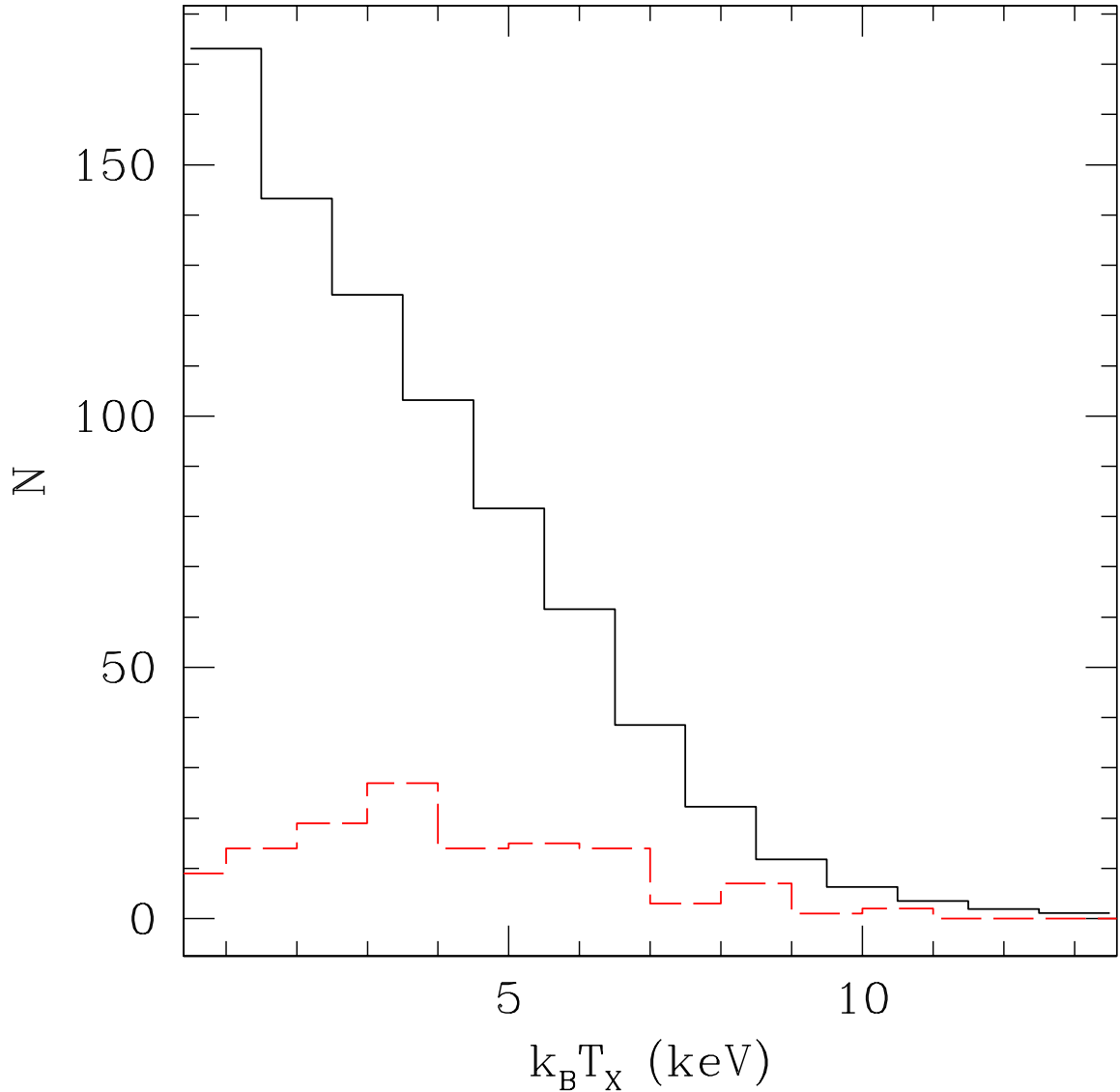}
\caption{\label{abundance} Expected cluster abundances in the nearby
($z\le 0.2$) universe over 4.24 str (upper histogram), compared to
the clusters used in our analysis (lower histogram). The expected
abundance is estimated for an X-ray survey with flux limit and sky
coverage similar to the REFLEX survey, using the concordant
cosmological parameters. The comparison suggests that there are many
more nearby clusters that can be added to our large sample, once
their temperature is measured. }
\end{figure}

    Turning to the future prospects, in the short term, the use of WMAP 4-year maps
    should decrease our errors on $f_{\rm gas}$ and concentration by a factor of $1.5-2$.
    Fig.(\ref{abundance}) shows the estimated number of
    observed clusters with $z<0.2$, within different temperature bins, which
    is compared to the sample with measured temperatures used for this
    analyses. This shows that, even within the range of temperatures
    and redshifts resolved by WMAP, there
    are many more clusters that can be included in the analyses, if
    their temperatures are measured. Therefore, adding more clusters with observed X-ray temperatures to
    our catalog, e.g., from current and future Chandra observations,
    will increase the significance of our constraints.

    In the long run, this work serves as an example to show the power of
    combining extragalactic observations in different frequencies in
    putting independent statistical constraints on the physical
    models of systems under study. More specifically, we have
    demonstrated that despite their low resolution, through combination with X-ray data, WMAP all-sky maps are
    capable of constraining the ICM physics at a comparable level of
    accuracy to their much higher-resolution pointed counterparts \citep[e.g., OVRO and BIMA; see][]{Grego2001}
    \footnote{However, note that the errors in \citet{Grego2001} is dominated by X-ray measurements, while the SZ detections are significant in
    each cluster. This is exactly opposite to the case analyzed in
    this paper.}
    Therefore, we predict that the scientific outcome of future high resolution CMB/SZ
    observations \citep[e.g., see][]{bond} will be greatly enhanced through direct
    combination/correlation with a wide-angle deep X-ray survey,
    such as NASA's proposed Dark Universe Observatory (DUO)
    \footnote{http://duo.gsfc.nasa.gov}. It is needless to say that
    more accurate observational constraints on the ICM need to be supplemented by
    more sophisticated theoretical models which should be achieved
    through large and high-resolution numerical simulations. Such simulations are yet to be realized (see \sec\ref{szft}).

\acknowledgments

We are grateful to Bruce Draine, Jim Gunn, Lyman Page, Matias
Zaldarriaga, and in particular to David Spergel for their comments
on the manuscript. The results in this paper have been derived using
the HEALPix \citep{healpix} package.

\appendix
\section{An Analytic Model of the Intra-Cluster Medium}

  Numerical simulations indicate that the spherically averaged density distribution of dark
   matter, which also dominates the gravitational potential of galaxy clusters, may be well approximated by an NFW
   profile \citep[][hereafter NFW]{NFW}
   \beq
   \rho(r) = \frac{\rho_s}{(r/r_s)(1+r/r_s)^2} ~{\rm , ~for}~ r < \rvir = \c
   ~r_s,\label{NFW}
   \eeq
   where $\rho_s$ and $r_s$ are constants, and $\c$, the so-called concentration
   parameter, is the ratio of $\rvir$ to
   $r_s$. The virial radius, $\rvir$, is defined as the boundary of the relaxed structure,
   generally assumed to be the radius of the sphere with an overdensity of $\Delta \simeq 200$
  with respect to the {\it critical density } of the universe
   \beq
    \left(4\pi\rvir^3\over 3\right) \left(\frac{3 H^2}{8\pi G}\right) \Delta = \mvir
    =\int^{\c r_s}_0 4\pi r^2 dr \cdot
    \rho(r) = 4 \pi \rho_s r^3_s \left[\ln (1+\c) -
% NOTE: minor change
%    \c/(1+\c)\right].\label{delta}
    {\c \over 1+\c } \right].\label{delta}
   \eeq
   Thus, fixing the mass of the cluster, $\mvir$, and the concentration
   parameter, $\c$, at a given redshift (which sets the critical density for given cosmology)
   fixes the dark matter profile ($\rho_s$ and $r_s$), and the associated gravitational potential
   \beq
   \phi(r) = - \frac{G \mvir}{r} \cdot
   \frac{\ln(1+r/r_s)}{\ln(1+\c)-\c/(1+\c)}.\label{phi}
   \eeq

    To model the distribution of the diffuse gas in the Intra-Cluster
    Medium (ICM), following \citet{Suto}, we assume that the gas follows a polytropic relation, i.e. $P_g(r) = \rho_g(r) T(r)/(\mu m_p) \propto
    \left[\rho_g(r)\right]^{\geff}$, and that it satisfies Hydrostatic
    equilibrium in the NFW potential, which reduces to
    \beq
    \frac{d}{dr}\left[ T(r)+ \mu m_p (1-\geff^{-1}) \phi(r) \right]
    =0. \label{hydro}
    \eeq
      Here, $P_g(r)$, $\rho_g(r)$, and $T(r)$ are gas pressure, density,
      and temperature respectively, while $\geff$ is the effective polytropic index of the
      gas and $m_p$ is the proton mass. $\mu = 4/(3+5X) \simeq 0.59$ is the mean molecular
      weight for a cosmic hydrogen abundance of $X\simeq 0.76$ .

      In order to integrate equation (\ref{hydro}), we need to set
      the boundary condition for $T(r)$. Assuming
      an accretion shock at $\rvir$ \citep{voit}, which causes the cold infalling gas
      to come to stop, the gas temperature behind the shock should
      be $T(\rvir) = \mu m_p v^2_{ac}/3$. In the spherical collapse
      model \citep{gunngott} $\rvir \simeq 0.5 \times r_{ta}$, the
      turn-around radius, which implies $v^2_{ac}=G \mvir/\rvir$, and respectively
      fixes $T(r)$
      \beq
       \frac{T(r)}{\mu m_p} \simeq - (1-\geff^{-1}) \left[\phi(r) -
       \phi(\rvir)\right] +\frac{G \mvir}{3 \rvir}.\label{temp}
       \eeq

      Note that, as our SZ signal is dominated by the most massive
      clusters (see \sec \ref{results}), it is a fair approximation to neglect
      the non-gravitational heating/cooling processes, which only
      become significant for smaller clusters \citep[e.g.,][]{voit}. Both simulations and
      observations seem to indicate that $\geff \simeq 1.2$
      \citep[e.g., FRB01,][and refernces therein]{voit,borgani}, and
      thus, for the rest of our analyses, we will use this value. Also,
      simulations only predict a weak mass dependence for the concentration parameter, $\c$
      \citep[e.g., NFW;][]{Eke2001,dolag}, and thus, for simplicity, we assume a mass-independent value of $\c$.
% NOTE: add one sentence
In \sec \ref{results} we explore the sensitivity of our observation
on $\c$.

      Combining equations (\ref{delta}),(\ref{phi}), and
      (\ref{temp}), we arrive at:
      \beq
      T(r) = \frac{\mu G \mvir m_p}{\rvir}~
      f(r/\rvir; \c, \geff),
      \eeq
      where
      \beq
      f(x; \c, \geff) = {1\over 3} +(1-\geff^{-1})\left[\frac{\ln(1+\c x)/x
      -\ln (1+\c)}{\ln(1+\c)-\c/(1+\c)}\right],\label{fdef}
      \eeq
        and
      \beq
      \rvir = \left(\frac{2 ~G \mvir}{H^2
      \Delta}\right)^{1/3}.\label{rvir}
      \eeq
       The polytropic relation can be used to obtain the ICM gas
       density, $\rho_g(r) \propto [T(r)]^{1/(\geff-1)}$, which
       yields
      \beq
      \rho_g(r) = f_{\rm gas} \left(\frac{\mvir}{4\pi \rvir^3}\right) g(r/\rvir; \c,
      \geff),\label{rhogas}
      \eeq
      where
      \beq
        g(x; \c, \geff) = \frac{[f(x; \c,
        \geff)]^{1/(\geff-1)}}{\int_0^1 [f(y; \c, \geff)]^{1/(\geff-1)}~ y^2
        dy},\label{gdef}
      \eeq
      and $f_{\rm gas}$ is the fraction of total mass in the ICM gas.

      Now, let us obtain the observable quantities that are relevant to our study.
      Similar to previous works \citep[e.g.,][]{Suto,komatsu}, we approximate the
      observable X-ray temperature of clusters as the X-ray emission
      weighted gas temperature
      \beq
      T_{\rm x} \simeq \frac{\int T^{3/2}(r) \rho^2_g(r) ~r^2 dr}{\int T^{1/2}(r) \rho^2_g(r) ~r^2
      dr} = \frac{\mu G \mvir m_p}{\rvir}~ \frac{\int^1_0 f^{3/2}(x)
      g^2(x)~ x^2 dx}{\int^1_0 f^{1/2}(x) g^2(x)~ x^2 dx},\label{tx}
      \eeq
       which should be contrasted with the virial (mass-weighted)
       temperature
      \beq
        \tvir = \frac{\int T(r) \rho_g(r) ~r^2 dr}{\int \rho_g(r) ~r^2
      dr} = \frac{\mu G \mvir m_p}{\rvir}~ \frac{\int^1_0 f(x)
      g(x)~ x^2 dx}{\int^1_0 g(x)~ x^2 dx}.\label{tvir}
      \eeq

        As an example, for a nominal value of $\c=3$, we find
      \beq
        T_{\rm x} = (8.7 ~\kev) \left(\frac{\mvir}{10^{15} h^{-1}
        \msun}\right)^{2/3}, {\rm ~ and ~}  \tvir = (7.2 ~\kev) \left(\frac{\mvir}{10^{15} h^{-1}
        \msun}\right)^{2/3},
      \eeq
        while, for $\c=5$
      \beq
        T_{\rm x} = (9.9 ~\kev) \left(\frac{\mvir}{10^{15} h^{-1}
        \msun}\right)^{2/3}, {\rm ~ and ~}  \tvir = (7.6 ~\kev) \left(\frac{\mvir}{10^{15} h^{-1}
        \msun}\right)^{2/3}.
      \eeq
        We note that
        these relations are consistent, at the $10\%$ level, with
        the predictions of the universal gas profile model by
        \citet{komatsu}. Observations of X-ray mass-temperature relation are often expressed
     in terms of $M_{500}$, i.e. the mass enclosed inside the sphere with $\Delta =500$ (see equation \ref{delta}).
        For $\c=3(5)$, Equation (\ref{delta}) gives $M_{500}/\mvir = M_{500}/M_{200} = 0.648(0.722)$ yielding
        \beq
            M_{500}(\c=3{\rm ~or~ 5}) = (4.0{\rm~or~}3.7) \times 10^{13}
            h_{70}^{-1}\msun \left(T_{\rm x} \over 5~
            \kev\right)^{3/2},
         \eeq

         corresponding to
         \beq
            M_{200}(\c=3{\rm ~or~ 5}) = (6.2{\rm~or~}5.1) \times 10^{13}
            h_{70}^{-1}\msun \left(T_{\rm x} \over 5~
            \kev\right)^{3/2},
         \eeq
         where $h_{70} \equiv h/0.7$. \citet{borgani} argue that, on average, the beta
       model polytropic ($\beta\gamma$) mass estimates overestimate the normalization of the observed M-T relation by about $30\%$ at $\Delta = 500$.
       Taking this into account, we notice that the normalization of
       the M-T relation in our model is consistent with the
       observations for hot clusters ($T_{\rm x} > 2-3 \kev$), at
       the $\sim 10\%$ level \citep[e.g., see Table 3 in ][]{arnaud}. Therefore, we will use Equation (\ref{tx}) to relate the
       observed X-ray temperatures of our clusters to their virial
       masses and radii. The expected systematic error in this conversion will
       be at the 10\% level in mass estimates (or 3\% in virial
       radii).

\end{document}